\documentclass[a4paper,11pt]{article}
\pdfoutput=1 
\usepackage{subcaption}
\usepackage{jcappub,aas_macros} 

\usepackage[T1]{fontenc} 
\usepackage{xcolor}

\newcommand{\fnlloc}{f_{NL}^{\mathrm{loc}}}

\newcommand{\bphi}{b_{\phi}}

\newcommand{\sigz}{\frac{\sigma_{z}}{1+z}}

\usepackage{hyperref}
\usepackage{fontawesome} 
\usepackage{diagbox}
\usepackage{booktabs,multirow,tabularx}
\newcommand{\makecell}[2][@{}c@{}]{\begin{tabular}{#1}#2\end{tabular}}

\title{\boldmath  
Learning to Concentrate: Multi-tracer Forecasts on Local Primordial Non-Gaussianity with Machine-Learned Bias
}
\author[a,b,1]{James M. Sullivan,\note{Corresponding author.} }
\author[b,c]{Tijan Prijon}
\author[a,b,d,e]{Uro\v{s} Seljak}

\affiliation[a]{Department of Astronomy, University of California, Berkeley, CA 94720, USA}
\affiliation[b]{Berkeley Center for Cosmological Physics, University of California, Berkeley, CA 94720, USA}
\affiliation[c]{Faculty of Mathematics and Physics, University of Ljubljana, Jadranska ulica 19, 1000 Ljubljana, Slovenia}
\affiliation[d]{Department of Physics, University of California, Berkeley, CA 94720, USA}
\affiliation[e]{Lawrence Berkeley National Laboratory, Berkeley, California 94720, USA}

\emailAdd{jmsullivan@berkeley.edu}
\emailAdd{tp7058@student.uni-lj.si}
\emailAdd{useljak@berkeley.edu}

\abstract{
Local primordial non-Gaussianity (LPNG) is predicted by many non-minimal models of inflation, and creates a scale-dependent contribution to the power spectrum of large-scale structure (LSS) tracers, whose amplitude is characterized by $\bphi$.
Knowledge of $\bphi$ for the observed tracer population is therefore crucial for learning about inflation from LSS.
Recently, it has been shown that the relationship between linear bias $b_1$ and $\bphi$ for simulated halos exhibits significant secondary dependence on halo concentration.
We leverage this fact to forecast multi-tracer constraints on $\fnlloc$. 
We train a machine learning model on observable properties of simulated IllustrisTNG galaxies to predict $\bphi$ for samples constructed to approximate DESI emission line galaxies (ELGs) and luminous red galaxies (LRGs). We find $\sigma(\fnlloc) = 2.3$, and $\sigma(\fnlloc) = 3.7$, respectively.
These forecasted errors are roughly factors of 3, and 35\% improvements over the single-tracer case for each sample, respectively.
When considering both ELGs and LRGs in their overlap region, we forecast $\sigma(\fnlloc) = 1.5$ is attainable with our learned model, more than a factor of 3 improvement over the 
single-tracer case, 
while the ideal split by $\bphi$ could reach $\sigma(\fnlloc) <1$. 
We also perform multi-tracer forecasts for upcoming spectroscopic surveys targeting LPNG (MegaMapper, SPHEREx) and show that splitting tracer samples by $\bphi$ can lead to an order-of-magnitude reduction in projected $\sigma(\fnlloc)$ for these surveys.
}

\begin{document}
\maketitle
\flushbottom

\section{Introduction}
\label{sec:intro}

Cosmological observations of the Cosmic Microwave Background (CMB) and Large-scale Structure (LSS) are currently in complete concordance with Gaussian initial conditions \cite{2019BAAS...51c.107M}.
Yet, the tantalizing possibility that there are deviations from Gaussianity lurking beneath observational uncertainties remains.
These deviations - or Primordial non-Gaussianity (PNG) - come in several well-known forms, one of which is the \textit{local} type (LPNG) parameterized by the amplitude $\fnlloc$:
\begin{equation}
    \phi = \phi_{G} + \fnlloc \left[\phi_{G}^{2} - \langle\phi_{G}^{2}\rangle\right],
    \label{eqn:bardeen}
\end{equation}
where $\phi$ is the Bardeen potential \cite{bardeen_potential} and $\phi_{G}$ a Gaussian random field.
A detection of $\fnlloc \sim 1$ would definitively indicate the presence of multiple fields during inflation, and provides a natural theoretical sensitivity target  \cite{2014arXiv1412.4671A, 2022arXiv220308128A}.

The Planck satellite has placed the most stringent constraints on $\fnlloc$ to date, finding $\fnlloc = -0.9 \pm 5.1$ \cite{PlanckPNG}, consistent with no local PNG.
While future CMB missions should improve on this uncertainty, they are not forecasted to reach $\sigma(\fnlloc)\lesssim 1$ \cite{cmbs4_sb,so_sb}. 
However, LSS surveys promise to supersede this sensitivity in the near-term \cite{dePutterDore17,Noah21, SchmittfullSeljak, Giri22, Dizgah21, 2022JCAP...04..013A, 2023arXiv230102406J}. 
In particular, the multi-tracer technique \cite{UrosMulti,2009JCAP...10..007M,simone_multi_smith,2011MNRAS.416.3009B,2023arXiv230102406J,2022JCAP...04..013A,2021PhRvD.104l3520D,2022JCAP...04..021M,2020MNRAS.498.3470W,2020RAA....20..158W} provides a powerful tool for using large-scale modes most affected by survey sample variance. 

The presence of local PNG modulates the halo density field in a scale-dependent manner \cite{Dalal08,Slosar08}.
This ``scale-dependent bias'' effect (LPNG bias) produces a signal in the halo power spectrum that scales like $k^{-2}$, becoming more important on larger scales.
This behavior is fundamentally due to LPNG inducing a bispectrum that peaks in the squeezed triangle configuration, which couples a long wavelength mode with two short wavelength modes, effectively linking very large scales with the small-scale process of halo formation.
The amplitude of this effect is controlled by the LPNG bias $\bphi$, and while initially it was believed that it only depends on the halo mass \cite{Dalal08}, it was soon realized secondary halo properties such as merger history also affect its value \cite{Slosar08}.
This effect has been exploited over the last 15 years in several LSS analyses to make significant progress toward a precision measurement of $\fnlloc$ \cite{Slosar08, LeistedtSDSSphoto, MuellereBOSS, CastoriaeBOSS, Cabass22BOSS, DAmico22BOSS}, though those constraints remain looser than those from Planck - LSS has yet to take the lead in constraining $\fnlloc$.
The above analyses do not make use of the multi-tracer technique, and therefore can in principle be pushed further in sensitivity.
However, many these analyses make a simplifying assumption that the relationship between the LPNG bias $\bphi$ and the linear tracer bias $b$ is perfectly known, or otherwise only consider a few values of constant shifts from this relationship \footnote{Though more recent work has also considered priors on $\bphi$ and constraining only $\bphi \fnlloc$  \cite{2022JCAP...11..013B,DAmico22BOSS,Cabass22BOSS} to account for this uncertainty}. 

A slew of recent papers has called this knowledge, which was always approximate, into significant question for simulated galaxies and other tracers \cite{2020JCAP...12..013B,2020JCAP...12..031B,2022JCAP...01..033B,2022JCAP...11..013B,marinucci_sham}.
Building on the pioneering work of Ref.~\cite{reid10_assmblyhistory}, Ref.~\cite{Lazeyras22} investigated to what extent several commonly considered assembly bias parameters affect the relationship between the linear halo bias $b_1$ and $\bphi$ in dark matter only N-body simulations.
For halo spin and sphericity, the authors found only a modest effect, but, especially for lower halo masses and redshifts, they found a significant secondary dependence of $\bphi$ on halo concentration.
There, the authors of Ref.~\cite{Lazeyras22} argue that the strong sensitivity of $\bphi$ to changes in concentration indicates that existing constraints assuming a fixed form for $\bphi(b_1)$ are unreliable, but also acknowledge that this sensitivity is an opportunity to better constrain $\fnlloc$ with suitably defined galaxy populations.
The goal of this work is to identify such populations with the aid of a machine learning algorithm trained on observable galaxy properties and to use its predictions to forecast the improvement on $\sigma(\fnlloc)$ that such a definition furnishes.

By using simulated galaxy samples that approximate the DESI selections for emission line galaxies (ELGs) and luminous red galaxies (LRGs), we can get a realistic estimate for the current and near-term prospects of constraining $\fnlloc$ by using observable quantities beyond halo mass to which $\bphi$ responds.
We will show that, depending on the information available and the galaxy sample properties, we are able to improve the forecasted error on $\fnlloc$ by factors of several for DESI-like galaxies using machine-learned $\bphi$ predictions.
We also comment on the prospects for using this strategy in future spectroscopic surveys, where naively one could expect an order of magnitude improvement on $\sigma(\fnlloc)$ if halo concentration were perfectly recoverable.

This paper is structured as follows:
We briefly review the LPNG parameter $\bphi$ in Section~\ref{sec:bphi_review} before moving on to the construction of the DESI-like simulated galaxy samples and our machine learning methodology in Section~\ref{sec:GNN}. 
We then provide forecasts for $\sigma(\fnlloc)$ in Section~\ref{sec:fisher}, and provide some concluding remarks in Section~\ref{sec:conclusions}.

\section{Local PNG bias $\bphi$}
\label{sec:bphi_review}

We briefly review the salient facts about the scale-dependent bias signal induced by local primordial non-Gaussianity and its amplitude $\bphi$.

In the peak-background split (PBS) formalism \cite{DJS}, the parameter $\bphi$ is defined as the response of the tracer mean density $\bar{n}_{t}$ to the presence of a long-wavelength perturbation of the Bardeen potential $\phi$ 
\begin{equation}
\bphi = \frac{\partial \log \bar{n}_{t}}{\partial(\fnlloc \phi)} = 2\frac{\partial \log \bar{n}_{t}}{\partial \log \sigma_8}.
\label{eqn:pbs}
\end{equation}
The second equality follows from the argument that the effect of LPNG on the tracer abundance is equivalent to the effect of a rescaling of the amplitude of linear fluctuations \cite{DJS}.
We will accept this argument for the purposes of this work.

This bias enters the tracer power spectrum at linear order in $\delta$ and $\fnlloc$:
\begin{align}
    P_{tt}(k) &= \langle \delta_{t}(\mathbf{k}) \delta_{t}^{*}(\mathbf{k}') \rangle' \\
    &\stackrel{L.O.}{=} b^{2} P_{L}(k) + 2\bphi b \mathcal{M}^{-1}(k) P_{L}(k),
    \label{eqn:bphi_power}
\end{align}
where the $'$ indicates that we drop the missing momentum-conserving Dirac delta and associated $(2\pi)^{3}$ prefactor, $b$ is the linear bias, and $\mathcal{M}^{-1}(k) = \frac{3\Omega_{m} H_{0}^{2}}{2 c^{2} D_{md} k^{2} \tilde{T}(k)}$, where $\tilde{T}(k)$ is the transfer function normalized to 1 on large scales, and $D_{md}$ is the linear total matter growth factor normalized to 1 in matter domination.

Reducing to the case of halos ($\bar{n}_{t} = \bar{n}_{h}$) with a universal mass function that only allows halo abundance to depend on halo mass, gives the universality relation (or UMF \cite{Slosar08}):
\begin{equation*}
\bphi(b,p) = 2 \delta_{c} (b-p),
\label{eqn:umf}
\end{equation*}
where the critical overdensity is $\delta_{c} = 1.686$.
For the case where we select all of the halos of a given mass the universality relation gives $p=1$.
For dark matter halos there are 
some deviations of this $b_{\phi}(b_{1})$ relationship observed \cite{Biagettibphi,2022JCAP...11..013B}.
It has also been argued that other values of $p$ are more appropriate for certain sub-populations of halos, such as for halos that have undergone a recent merger, which may host quasars \cite{Slosar08}.
However, the picture becomes even more complicated for simulated galaxies, for which it was recently shown that when selecting by stellar mass, a value of $p=0.55$ is preferred when using the IllustrisTNG galaxy formation model \cite{2020JCAP...12..013B}.
Presumably, if the stellar mass selection can be replaced with halo mass selection, this would restore $p=1$, which suggests that by choosing additional observables one may be able to split galaxies by $\bphi$. 

One such additional observable that 
has been found to be very 
sensitive to $\bphi$
is halo concentration $c$. 
With a fixed galaxy formation model, a value of the LPNG bias can be estimated from Separate Universe (SU) simulations 
\citep[e.g.,][]{baldauf_su,sirko_su,gnedin_su,wagner_su,baldauf_su1,yin_su_14,pat_su,liang_su} in which two N-body simulations are run with two values of $\sigma_{8}$ and the bias is then estimated via finite difference with respect to the mean number density $\bar{n}_{h}(M,c)$. 
We use the reported results of Refs.~\cite{Lazeyras22,2020JCAP...12..013B} at $z=1$ to relate linear halo bias $b_{1}$ (estimated from power spectra) and SU LPNG bias $\bphi$ to halo mass $M$ and concentration $c$. 

\section{$\bphi$ with machine learning}
\label{sec:GNN}

In this Section, we describe our machine learning method for obtaining predictions for $\bphi$.
Before diving into the details, we provide a brief overview of this process here.

First, we obtain $b(M,c), b_{\phi}(M,c)$ by interpolating the gravity-only Separate Universe (SU) results of \cite{Lazeyras22} in halo mass $M$ and concentration $c$, and assign $b(M,c), b_{\phi}(M,c)$ to each TNG galaxy's host halo of mass $M$ and concentration $c$ .
We then train the ML model described below on galaxies where the values of $b_{\phi}$ are known to learn the relationship between input features such as galaxy magnitude and $\bphi$. 
After making a prediction for $\bphi$ on test data, we rank order the $\bphi$ predictions to determine a split of the sample into tertiles.
We then obtain predictions for a mean value of $\bphi$ in each tertile by using the trained ML model evaluated on the observable input features of galaxies in that tertile (magnitude, position, etc.) and averaging.
This final set of $\bphi$ values in the tertiles are used as input to the Fisher forecasts of Section~\ref{sec:fisher}.

\subsection{Simulated galaxy sample}

\subsubsection{Simulation}

\begin{figure}[h!]
    \centering
    \includegraphics[width=0.7\textwidth]{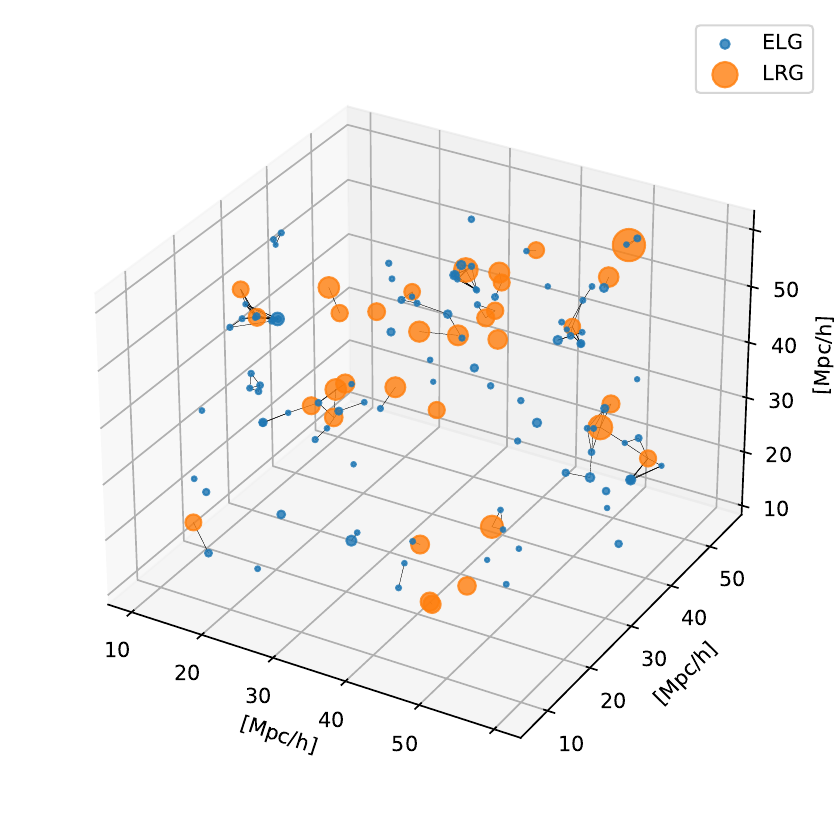}
    \caption{ELG and LRG-like galaxies in the part of the volume of the IllustrisTNG simulations at redshift $z=1$. Galaxies are presented in real space, and those closer than $5\ h^{-1}~\mathrm{Mpc}$ are linked for ease of presentation. The size of each galaxy marker is directly proportional to its logarithmic stellar mass ($\log{M^*}$).}
    \label{fig:simulation}
\end{figure}

We use simulated galaxy data from the IllustrisTNG-300 \cite{Nelson2018-wc,Pillepich2017-nl,Springel2017-pb,Nelson2017-th,Naiman2017-mv,Marinacci2017-fu} simulation, a cosmological magnetohydrodynamical simulation that models the formation and evolution of galaxies and covers the volume of $V = 205^{3}\ [h^{-1}~\mathrm{Mpc}]^3$. The snapshot that we use is at redshift $z=1$, which is a compromise between being able to mimic Dark Energy Spectroscopic Instrument (DESI) survey \cite{desi2016doc} observations of both emission-line galaxies (ELGs) and luminous red galaxies (LRGs). A sub-volume of the simulation containing ELG and LRG- like galaxies, as we define in Section \ref{sec:sample}, is shown in Fig. \ref{fig:simulation}.

\subsubsection{Observables}\label{sec:observables}
The halos in the IllustrisTNG simulation are determined with the friend-of-friends (FoF) algorithm \cite{davis_fof}, with linking length $\ell=0.2$.

We use the results of Ref. \cite{Lazeyras22} to link $\bphi$ and $b$ to halo mass and concentration. We take halo concentration from the IllustrisTNG supplementary data catalog \cite{Anbajagane2021}. It is defined as $c=R_{200c}/R_s$, where $R_s$ is determined by fitting a Navarro-Frenk-White (NFW) \cite{nfw} profile to the dark matter density profile. 
We use the $R_{200c}$ halo mass definition.
We note that Ref.~\cite{Lazeyras22} used several halo mass definitions when using different simulations, though since concentration was logarithmized, the effect of these differences should be reduced.

To avoid using halo properties that are unobservable by spectroscopic surveys to infer halo LPNG bias, we only use stellar masses $M^*$, the $r$, $g$, and $z$ magnitudes, and redshift-space positions of the galaxies. 
We use the word ``observable'' somewhat loosely, as galaxy stellar mass must be estimated from spectra, and dust models that affect the observed $r,g,z$ magnitudes are uncertain - we do not account for these aspects that would be relevant for a real data analysis here.
We choose the IllustrisTNG $z$-coordinate as the line-of-sight direction to obtain positions in redshift space.
The real-to-redshift space transformation is 
\begin{equation}
    \mathbf{s} = \mathbf{x} + \frac{v(\mathbf{x})\mu}{\mathcal{H}} \hat{z}
    \label{eqn:rsd}
\end{equation}
where $\mathbf{x}$ is the real-space galaxy position, $v(x)$ is the galaxy velocity magnitude, $\hat{z}$ is the line-of-sight direction, $\mu = \hat{\mathbf{v}}\cdot\hat{z}$, $\mathcal{H}$ is the conformal Hubble parameter, and $\mathbf{s}$ is the redshift-space position.
To simulate the effect of redshift-space distortions on galaxy observations, we transform the galaxies to redshift space before computing input features and training the machine learning algorithm to predict $\bphi$. 

While each galaxy lies in exactly one halo, a halo can contain multiple galaxies inside. When we report results in later sections, we predict $\bphi$ of the halo for each galaxy separately. Thus, it can happen that in the results, we have several (either the same or different) predictions for the same halo if several galaxies lie in it. This approach is justified because we, in reality, observe galaxies and cannot a priori know whether two galaxies belong to the same halo. We emphasize that in the training we do not use any information about the halos (for example, halo position), except the knowledge of to which parent halo a galaxy belongs.

\subsubsection{ELG and LRG sample}\label{sec:sample}
Our data sample mimics the emission-line galaxies (ELGs), and luminous red galaxies (LRGs), as observed by the DESI survey at redshift $z=1$. We can construct these mock galaxy samples either by color-space cuts or by galaxy selection based on specific star-formation rate (sSFR, i.e., the SFR per stellar mass) and stellar mass ($M^*$) cuts \cite{Hadzhiyska2021, Yuan2022-wi, 2210.10068}. We adopt the latter strategy and use sSFR and stellar mass cuts.

\begin{figure}[h!]
    \centering
    \includegraphics[width=1.\textwidth]{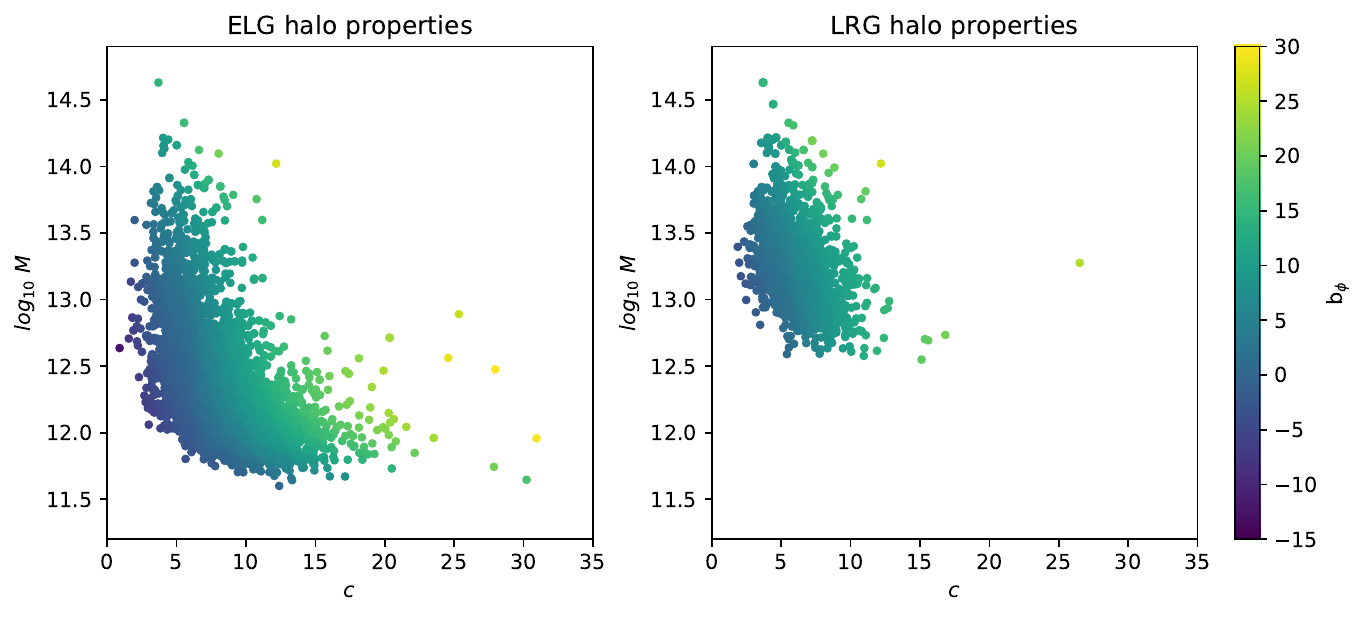}
    \caption{Parent halo properties of ELGs and LRGs. As expected we see that LRGs live in more massive halos with lower concentration. We also show the $\bphi$ value associated each halo. We can see that while it is a function of both halo mass and concentration, the latter has a much greater impact. The halo mass is reported in the units of $10^{10}\ \mathrm{M}_{\odot}/h$.}
    \label{fig:elg_lrg}
\end{figure}

To obtain both ELG and LRG samples, we first make sSFR cuts and then take the heaviest galaxies based on $M^*$ to get the desired target number densities. 
We made two different sample selections using sSFR cut values of $\log_{10}[sSFR] = -9.09$ from \cite{2210.10068}, and $\log_{10}[sSFR] = -9.23$ from \cite{Hadzhiyska2021}. 
The target number densities are taken from \cite{Zhou2023, AnandELG}. 
The selection of ELG and LRG samples is summarized in Table \ref{tab:sample_selection}. 
Here we find that the mean $b_1$ values of our mock samples are similar to those of Ref.~\cite{desi2016doc} for ELGs (1.4) and LRGs (2.6) at $z=1$.
With this selection, we also ensure that no galaxy is classified as an ELG and an LRG simultaneously.
In the next sections we report the results for all eight galaxy samples presented here. We optimize hyperparameter separately for each sample (see Section~\ref{sec:training}). 
However, because the samples are relatively similar, most hyperparameters remain the same.

\begin{table}[h!]
\centering
\begin{tabular}{|l|l|l|l|l|}
\hline
Type & $\log_{10}{sSFR}$ & $n$ $[h^{-1}~\mathrm{Mpc}]^{-3}$& $\min{\log_{10}{M^*}}$ & $\bar{b}$\\ \hline
LRG         & < -9.09    & $2 \times 10^{-4}$& 1.06 & 2.40      \\
LRG         & < -9.23    & $2 \times 10^{-4}$     & 1.06 & 2.40     \\
ELG         & > -9.09    & $5 \times 10^{-4}$    & 0.02 & 1.44      \\
ELG         & > -9.09    & $7 \times 10^{-4}$   & -0.09 & 1.42      \\
ELG         & > -9.09    & $1 \times 10^{-3}$    & -0.23 & 1.39     \\
ELG         & > -9.23    & $5 \times 10^{-4}$     & 0.23 & 1.46     \\
ELG         & > -9.23    & $7 \times 10^{-4}$     & 0.17 & 1.44     \\
ELG         & > -9.23    & $1 \times 10^{-3}$     & 0.07 & 1.42     \\ \hline
\end{tabular}
\caption{Selection criteria with number densities, minimum obtained stellar mass $M^*$ (units of $10^{10}\ M_{\odot} h$), and average $b_1$ for each sample. We first apply the sSFR cut (units of $h/\mathrm{yr}$), as specified in the second column, and then select galaxies with the highest $M^*$ to achieve the target number density. The number of galaxies (``size of dataset'') in each sample can be calculated from number densities and the volume of the simulation ($V = 205^{3}\ [h^{-1}~\mathrm{Mpc}]^3$), and lie between 1500 and 7000.}
\label{tab:sample_selection}
\end{table}

\begin{figure}[h!]
    \centering
    \includegraphics[width=0.7\textwidth]{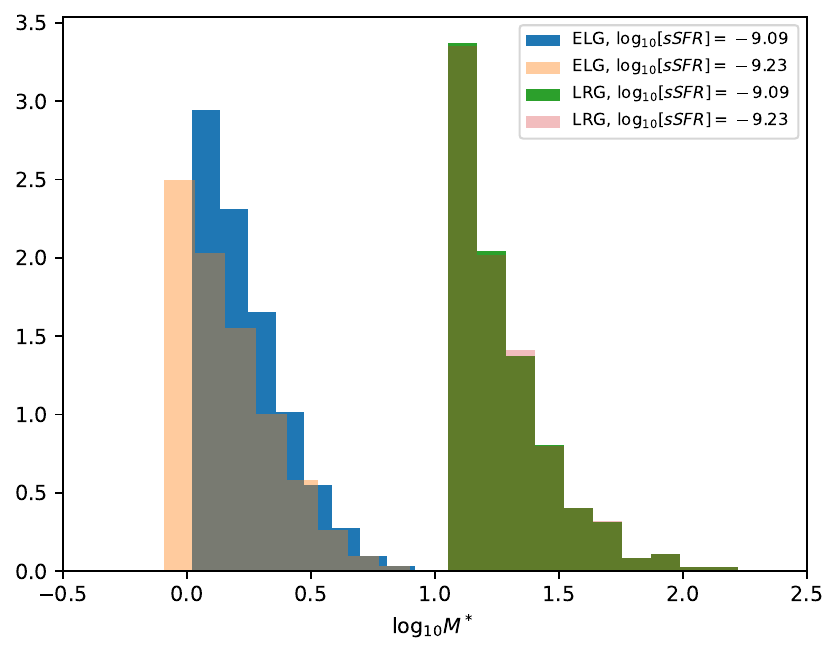}
    \caption{The stellar mass of ELG and LRG-like galaxies selected by two different sSFR cuts. LRGs are significantly heavier in both cuts, and it is apparent that the ELG and LRG selected samples do not overlap. While the different sSFR cuts significantly affect the $M^*$ of the ELGs, they have almost no influence on the $M^*$ of LRGs. The histograms are normalized for easier visualization, as ELG and LRG samples do not contain the same number of galaxies.}
    \label{fig:ssfr_cut}
\end{figure}

In Fig. \ref{fig:elg_lrg}, we compare the halo properties in which LRG and ELG galaxies lie, and show their expected $\bphi$ in the mass and concentration plane. We see an apparent trend that LRGs lie in heavier halos and that their halos have lower concentrations.
In Fig. \ref{fig:ssfr_cut}, we compare the stellar mass of ELGs and LRGs and its difference when applying different sSFR cuts. While the difference in sSFR affects only the $M^*$ distribution of ELGs, it is evident that LRGs are significantly heavier, in accordance with Refs. \cite{desi2016doc,rongpu_photo_lrg}.

All the results in this Section are reported for the samples selected with the sSFR cut of $\log_{10}{sSFR} = -9.09$, and with number densities of $n_{ELG}=5 \times 10^{-4}\ [h^{-1}~\mathrm{Mpc}]^{-3}$, and $n_{LRG}=2 \times 10^{-4}\ [h^{-1}~\mathrm{Mpc}]^{-3}$, as these are the DESI number densities at $z=1$ in Refs.~\cite{AnandELG,Zhou2023}.

\subsection{Machine learning}

Our objective is to predict the value of the continuous variable $\bphi$, based on $M$ continuous features (inputs)  $\mathbf{x_i} = (x_{i,1}, x_{i,2}, ..., x_{i,M})$ on $N$ training examples, which are in our case galaxies. We want to find a function $f: \mathbb{R}^M \rightarrow \mathbb{R}$ that maps the inputs $\mathbf{x_i}$ to the scalar value of $\bphi$. When constructing such a function, we want to minimize the difference between its output and the actual value of $\bphi$. The metric of the difference we are minimizing is our loss function, which we take to be the most common root-mean-square error function (RMSE), defined as $L = \sqrt{\frac{1}{N} \sum_{i=1}^{N} (y_i - \hat{y_i})^2}$, where $y_i$ and $\hat{y_i}$ are the actual and predicted values of $\bphi$ for the $i^{th}$ training example.

The described problem is a single-target regression problem. However, rather than first interpolating $\bphi$ from halo mass and concentration (as described in Section \ref{sec:observables}) and directly predicting it, we could make predictions for mass and concentration, and interpolate $\bphi$ from predicted values instead. In this case, we would be dealing with a multi-target regression problem. We tried this approach and verified that it leads to worse results according to the RMSE metric. The difference between the two approaches can be imagined only as a difference in the loss function that is optimized. However, because we are only interested in $\bphi$ it is both simpler and more effective to focus on a single target regression.

The problem of inferring the halo properties from observable data has already been studied in the literature using various machine learning (ML) techniques. One approach is to use structured data with a combination of graph neural networks (GNN) or convolutional neural networks (CNN), while another approach uses unstructured data by calculating statistics of the halo neighborhood \citep[e.g.,][]{VillanuevaDomingo2022, 2018MNRAS.478.3410A,2021MNRAS.507.2115M,2021MNRAS.507.4879X,2022MNRAS.515.2733D,2022MNRAS.514.2463D,2022MNRAS.514.4026S}. We focus on the latter and used several different ML algorithms in combination with unstructured input features. The main reason for this is that GNN and CNN would generally need a larger data sample size, with higher number densities. We further justify this selection when we discuss the interpretability of our model in Section~\ref{sec:results}.

\subsubsection{Input features}\label{sec:features}
Each galaxy belongs to exactly one halo whose $\bphi$ we want to infer. However, it is not necessarily true that the halo's center coincides with this galaxy's position. We define a central galaxy as the galaxy with the largest $M^*$ within $R=1\ h^{-1}~\mathrm{Mpc}$ of the galaxy whose halo properties we want to infer. $R$ is a hyperparameter of our model, which we optimize as discussed in Section \ref{sec:training}. We then use the central galaxy surroundings to describe the halo environment more accurately.

We use 13 features that we found useful for predicting $\bphi$: 
\begin{itemize}
\item $r, g, z$ -- AB magnitudes of $r$, $g$, and $z$ bands of the central galaxy
\item $\log{M^*}$ of the central galaxy
\item $N_{R}$;  number of neighbor galaxies within $R$, for $R \in \{0.5,1,2,3,4,5\}\ h^{-1}~\mathrm{Mpc}$
\item Metric of anisotropy, defined as 
$A = \left\Vert \sum_{i} \frac{\mathbf{R_i}}{\left\Vert \mathbf{R_i}\right\Vert} \cdot H(R_H - R_{i}) \right\Vert_2$
\item Sum of stellar masses of the galaxies in the neighborhood; $\log{\left(\sum_{i} M^*_{i} \cdot H(R_H - R_{i})\right)}$
\item Sum of stellar mass to distances ratios; $\log{\left(\sum_{i} \frac{M^*_{i}}{R_{i}} \cdot H(R_H - R_{i})\right)}$. 
\end{itemize}
Here $\mathbf{R_i}$ is the separation vector between the $i$-th neighbor galaxy and central galaxy in redshift space, and $H(R_H - R_i)$ is the Heaviside step function that ensures that only galaxies within the radius $R_H$ are included in the sum. $R_H$ is a hyperparameter of our model, for which we use $R_H = 5\ h^{-1}~\mathrm{Mpc}$. The results are mostly insensitive to the choice of $R_h$, however when considering a neighborhood much more distant than $5\ h^{-1}~\mathrm{Mpc}$, the results do not improve, and rather start to deteriorate. The machine learning models can in theory learn to ignore less important input features, however, we are limited by the size of our dataset. For this reason, using fewer, only dominantly important features turns out to be better. All the described features use the distances calculated in redshift space. The predictive power of the model would increase if the features are calculated in real space, however we do not focus on them, since spectroscopic surveys observe galaxies in redshift space. 

\begin{figure}[h!]
    \centering
    \includegraphics[width=1.\textwidth]{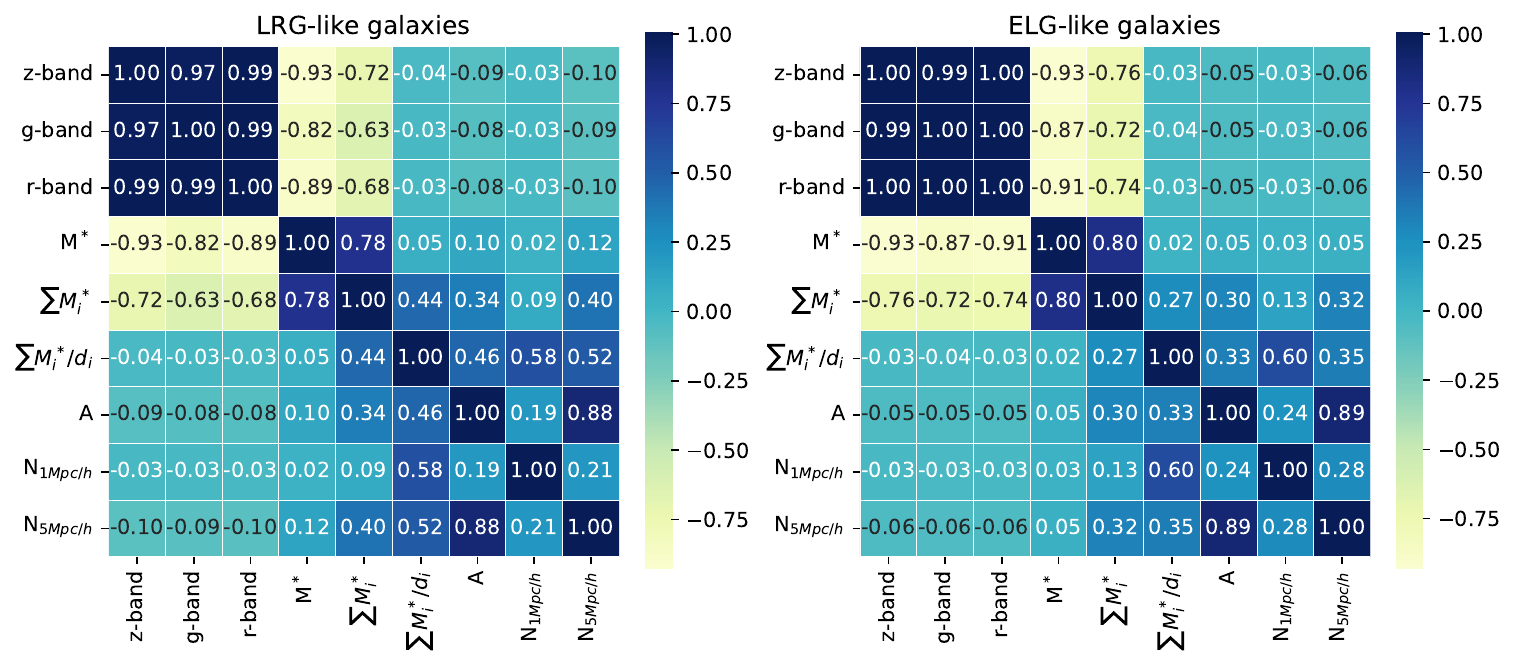}
    \caption{Pearson correlation coefficients of the most representative input features. The coefficients are rounded and are not exactly 1 between different luminosity bands. While the intrinsic features of galaxies (luminosity bands and stellar mass) are quite strongly correlated, they are mostly uncorrelated with the distributions of the galaxies in the neighborhood.}
    \label{fig:correlation}
\end{figure}

The correlations between the most representative input features are presented in Fig. \ref{fig:correlation}. The intrinsic features of the galaxies are strongly correlated, with the correlation between luminosity bands being almost exactly 1. For this reason, we tried performing principal component analysis (PCA), which would reduce the dimensionality of the dataset. However, the results did not improve, and we do not use PCA to produce final predictions.

The stellar mass of the central galaxy, sum of neighborhood stellar masses, and sum of stellar-mass-to-distance ratios are all logarithmized before being used for machine learning. They are approximately exponentially distributed, therefore very large (small) numerical values could lead to potential numerical instabilities during the ML learning process. Furthermore, we standardize all features, $\hat{x_i} = \frac{x_i - \mu_i}{\sigma_i}$. Normalizing instead of standardizing features lead to slightly worse results, probably due to the outliers present.

We experimented with utilizing alternative input features. We already commented on the rationale behind restricting the input information to neighboring galaxies within a radius of $R_H$. However, it would be possible to instead select the N nearest galaxies with the objective of optimizing N. This approach appears to be less intuitive from a physical standpoint, as it may not accurately capture the influence of clustering effects. We have indeed checked and observed that it leads to slightly worse predictions for $\bphi$. Furthermore, we have checked and observed that incorporating information on the luminosities and stellar masses of the neighboring galaxies does not lead to any discernible improvements in the obtained predictions. 

In Table \ref{tab:sample_selection}, we have 8 different datasets for which we want to make separate $\bphi$ predictions. When we make predictions for ELG galaxies, we also include information about nearby LRG galaxies (in redshift-space) and vice versa. 
In other words, we use both ELGs and LRGs to calculate the number of neighbor galaxies or local anisotropy. 
However, we train the model and make predictions only on one type of galaxy at a time - this procedure could be performed on real data.

\subsubsection{Models}\label{sec:models}
We have tried using several different machine learning algorithms to predict $\bphi$ from the input features as described in Section \ref{sec:features}. The top three best-performing models are Artificial Neural Networks (ANN), Support Vector Machines (SVMs), and gradient-boosted tree models (XGBoost \cite{xgboost}). We discuss the process of hyperparameter optimization in Section \ref{sec:training}.

We employ a technique known as model stacking to enhance the predictive performance of the three best-performing models; ANN, SVM, and XGBoost. Model stacking integrates individual predictions of multiple models and produces the final output from them. By stacking models, we can reduce both prediction bias and variance. The prediction errors arising from the individual models stem from distinct sources and thus can be partially offset.

The effectiveness of model stacking relies heavily on the diversity and quality of the base learners 
(the models whose predictions we are combining). A decrease in variance and new sources of prediction bias may be introduced if the base models are not good enough. ANN, XGBoost, and SVM all perform well independently and have distinct modeling techniques and assumptions (XGBoost is an example of a tree-based ensemble method, ANN is a type of neural network, and SVM is a discriminative model). We tried including additional models in the ensemble, which did not lead to an increase in predictive power.

After obtaining the predictions of the base models, we want to train a new meta-model on them and map ``base predictions'' to $\bphi$. We want to find a function $f: \mathbb{R}^{3} \rightarrow \mathbb{R}$, which takes three continuous inputs (predictions of base learners) and outputs $\bphi$. We use a simple ridge regression for this task to avoid overfitting.

\subsubsection{Training}\label{sec:training}
To optimize the hyperparameters of the algorithms described in Section \ref{sec:models}, and features described in Section \ref{sec:features}, we perform a train-validation-test split of the dataset (of each galaxy sample). The test set is held aside and is only used for the final evaluation. We use 3-fold cross-validation for training the meta-model.

Regarding the size of the train-test-validation split, we use a ratio of 70-20-10 \%. Because we are dealing with spatially distributed galaxies, having a random split would lead to data leakage. The galaxies in the test (or validation) set may lie nearby the galaxies from the training set and would thus be in an identical environment. It would allow the model to perform well on the test set by simply memorizing the patterns from the training set, but would not be able to generalize well. We adopt the solution of splitting the data into spatially distinct regions based on $x$ and $y$ spatial coordinates, which are not the line-of-sight directions.

Using the training and validation sets, we employ grid search to tune the hyperparameters. We also evaluate the performance of other combinations of input features, different options of feature transformations, different loss functions, different models, and the performance of adopting a multi-target regression problem and predicting halo mass and concentration over $\bphi$.

Examples of features that we optimized have already been mentioned in the section \ref{sec:features}. The optimizations include adjusting the distances to capture information about the surroundings and modifying the number of features that describe environmental density.

While optimizing the models, we explored a range of parameters. However, the optimizations improved the results only to some extent because the correlation between the input features we use and the halo assembly bias is limited (in addition to being complex). 

Regarding the artificial neural networks (ANN), we have tried a range of hidden layers between 2 and 10, the number of neutrons per layer between 5 and 50, and we also changed the dropout rate and the activation functions. 
Our final ANN architecture has four hidden layers with 20 neurons per layer. In addition, we use dropout and the LeakyReLU activation function. The kernel of the SVM model used is the radial basis function. All other values of the hyperparameters and the values in the grid search grid are available on GitHub\footnote{\url{https://github.com/jmsull/ml_fnl_forecast}}.

\subsection{Results}\label{sec:results}

\begin{figure}[h!]
    \centering
    \includegraphics[width=1.\textwidth]{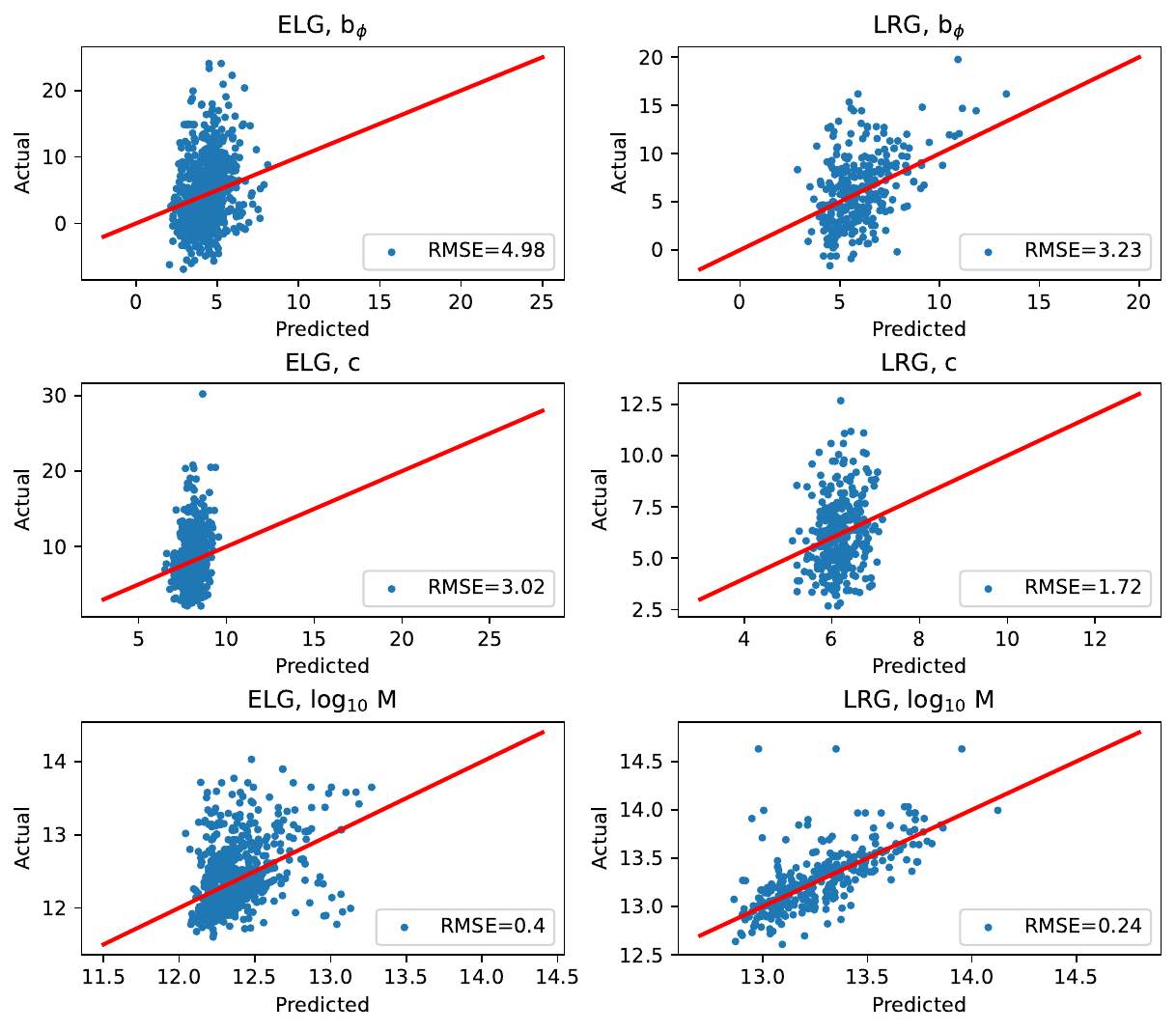}
    \caption{Results for ELG (left) and LRG (right) predictions on the test set. We present predicted versus actual values for each galaxy. The perfect predictions should lie on the red line, which has a slope of 1. Only the $\bphi$ predictions (top row) are relevant for the analysis. However, we also show halo concentration and mass predictions (bottom two rows).}
    \label{fig:results}
\end{figure}

The final predictions made by the stacked model on the test set are presented in Fig. \ref{fig:results}.. We also show the separate predictions for halo concentration and mass. We do not use mass and concentration predictions elsewhere, but it is useful to see them since $\bphi$ is a function of these variables. We can see that for both ELGs and LRGs, halo mass predictions are significantly better than both $\bphi$ and concentration predictions. While the $\bphi$ and concentration RSME are of the same magnitude, the RMSE of mass prediction is roughly 10 times smaller. 

It is hard to make direct comparison between ELG and LRG results, since the distributions of the target values are different, as seen in Fig. \ref{fig:elg_lrg}. However, it seems that the results in terms of RMSE are generally better for LRGs. The reason for this could be that, as seen in Fig. \ref{fig:elg_lrg}, LRGs live in more massive halos, which are correlated with larger environments and higher central fractions. Moreover, LRGs are older than ELGs and are thus less affected by hydrodynamical effects, which are hard to capture with the properties we are looking at.

\begin{figure}[h!]
    \centering
    \includegraphics[width=1.\textwidth]{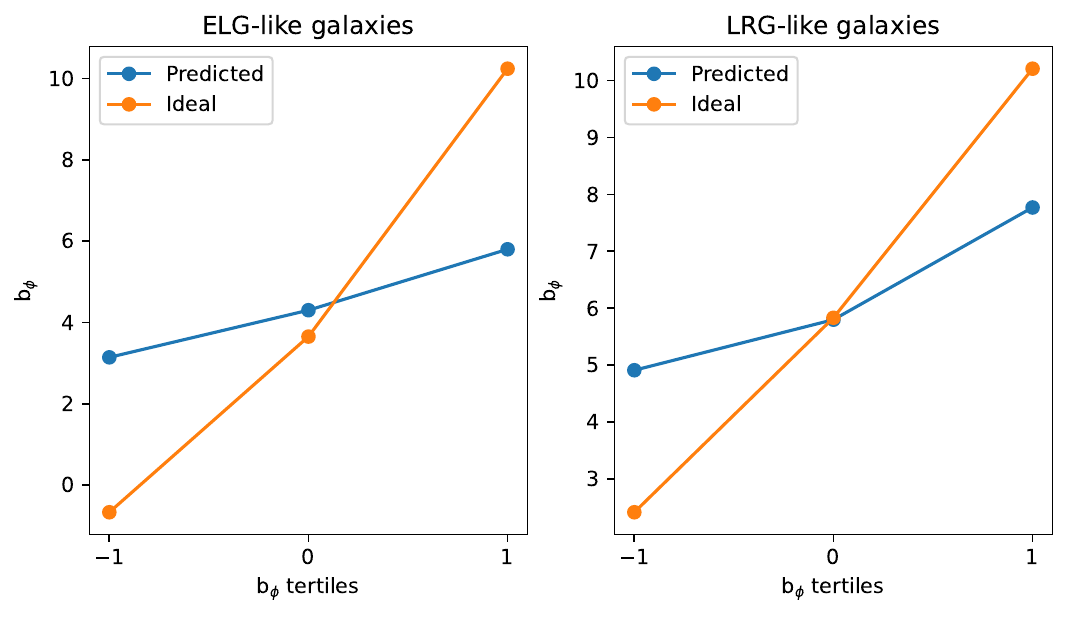}
    \caption{$\bar{b}_{\phi}$, for ELG and LRG samples. We split the sample into tertiles for the ``Ideal'' case based on the actual $\bphi$ value and report the average $\bphi$ for each tertile. For the ``Predicted'' case, we make the split based on predicted $\bphi$, as shown in Fig. \ref{fig:results}. The horizontal axis on the plots is arbitrary and just denotes different tertiles.}
    \label{fig:bphi_results}
\end{figure}

In Fig. \ref{fig:bphi_results} we can see the average $\bphi$ values of the tertiles, split based on predicted $\bphi$ from Fig. \ref{fig:results}. The ``Ideal'' dots in Fig. \ref{fig:bphi_results} correspond to the values of $\bphi$ for splits made by actual $\bphi$ values of the galaxy host halos. After these tertilles are defined we compute the mean $b_1$ of all halos in each tertile.

\begin{figure}[h!]
  \centering
  \begin{subfigure}[b]{0.4\textwidth}
    \includegraphics[width=\textwidth]{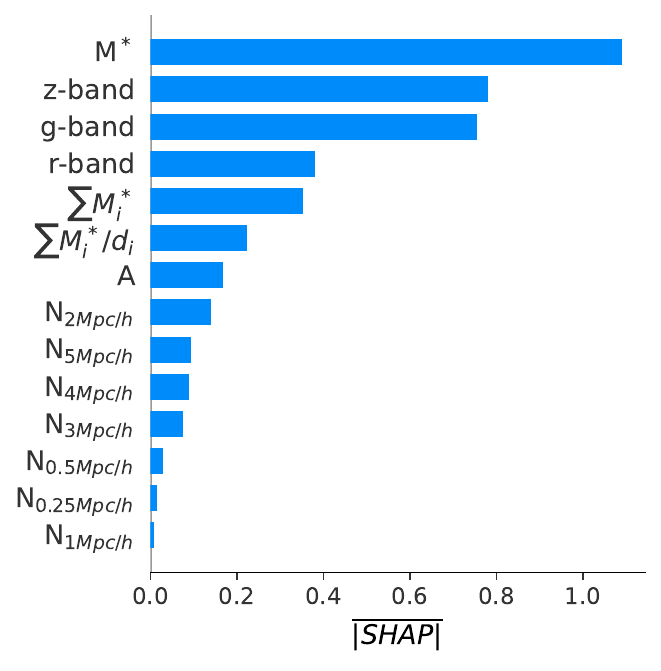}
    \caption{ELG-like galaxies}
    \label{fig:ELG_importance}
  \end{subfigure}
  \hfill
  \begin{subfigure}[b]{0.4\textwidth}
    \includegraphics[width=\textwidth]{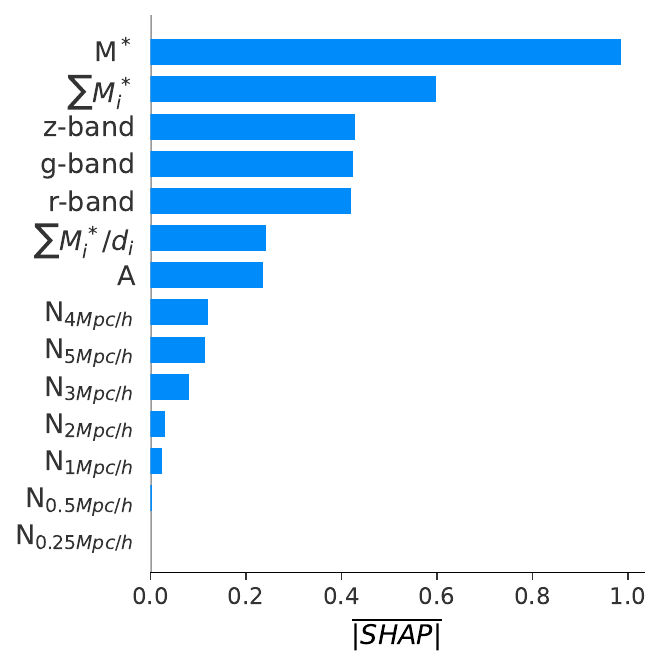}
    \caption{LRG-like galaxies}
    \label{fig:LRG_importance}
  \end{subfigure}
  \caption{SHAP values of feature importance for predictions in redshift space. We can see that for both ELG and LRG- like galaxies, intrinsic features of galaxies are of much greater importance than the properties of the neighborhood. We can also observe that the feature importance is similar for both ELGs and LRGs.}
  \label{fig:feature_importance}
\end{figure}

We want to look at the interpretability of our model and see which information is most important for the obtained results. For this purpose, we use Shapley Additive explanations (SHAP) values, a technique for determining feature importance \cite{lundberg2017unified}. The SHAP feature importance is calculated by permutating all features and summing the average contribution of each feature over all permutations. It provides a fair and accurate estimation of the contribution of each feature, even if features are correlated, and can account for complex interactions between features\footnote{The more popular technique used is ``permutation importance'', which has the main drawback of being unable to deal with correlated features.}.

The average SHAP values for each feature are presented in Fig. \ref{fig:feature_importance}. The feature importance is presented for only two galaxy samples, however it remains practically the same for others. We can notice that, for both ELG and LRG, the intrinsic properties of the galaxies are significantly more important than environmental properties. Another interesting thing to observe is that the feature importance is similar for ELG and LRG galaxies.

We can observe that the importance of $N_{R}$ does not converge to 0. This makes sense since there is a high correlation between the number of neighbors within different $R$. Moreover, we have explicitly checked that adding information about a larger environment does not improve results. It just takes some importance from other $N_R$ features, while their sum would remain the same. The reasons are that the information too far from the halo does not affect the results, the dataset size is inadequate, and the correlation between halo properties and those features is noisy.

In Fig. \ref{fig:feature_importance}, we can see that the information from the local neighborhood is much less influential than other features are when making predictions. Furthermore, we observed that adding information about luminosities and stellar masses of nearby galaxies does not improve the results. We can conclude that it thus makes sense to approach the problem with the unstructured data instead of using GNN or CNN. Those could work much better in real space, with a larger simulation (larger data size) and higher number densities.

\section{Multi-tracer $\fnlloc$ forecasts}
\label{sec:fisher}

In this Section we provide Fisher forecast results for $\sigma(\fnlloc)$ using a slightly extended version of the multi-tracer approach of Ref.~\cite{UrosMulti}.
In particular, we first provide forecasts for DESI LRGs and ELGs using the $\bphi(b_{1})$ relations found in Section~\ref{sec:GNN} for mock versions of these samples.
We then provide more speculative forecasts for future planned spectroscopic surveys (MegaMapper \cite{simone_highz_spectro}, SPHEREx \cite{spherex14}) that will target LPNG.

\subsection{Forecast setup}
\label{subsec:fisher_setup}
Following \cite{UrosMulti}, we perform forecasts of idealized constraints on $\fnlloc$, slightly generalizing the expressions there to a 3-tracer analysis.
To be consistent with IllustrisTNG, we will use the Planck15 fiducial flat $\Lambda$CDM cosmology with $\Omega_m = 0.3089$, $\Omega_b = 0.0486$,  $\Omega_\Lambda = 0.6911$, $h = 0.6774$, $\sigma_8 = 0.8159$, and $n_{s} = 0.9667$ in these forecasts \cite{Planck15,TNGoriginal}, and do not marginalize over $\Lambda$CDM parameters.
While this choice might be considered optimistic, the multi-tracer method is relatively insensitive to marginalizing over $\Lambda$CDM parameters (see Appendix C of Ref.~\cite{hamaus_halo_forecast}).
Furthermore, even without the multi-tracer method, we find that our single-tracer forecasts produce $\sigma(\fnlloc)$ that is 10\% lower than the (marginalized) values quoted in  Ref.~\cite{Noah21} or less (see Fig. 14 there) except for DESI ELGs, where the difference is 25\%, though in this case, we do not use the same redshift distribution or redshift range as in Ref.~\cite{Noah21}.
We will work entirely in linear theory at the power spectrum level (including linear redshift space distortions $b \to b + f\mu^{2}$), and use a simple linear bias model only.
We do not account for the Alcock-Paczynski effect \cite{1979Natur.281..358A}.
We compute the linear matter power spectrum using \texttt{camb} \cite{camb}.
We use $101$ linear $\mu$ bins between 0 and 1, $N_{z}=65$ linear $z$ bins between $z_{\mathrm{min}}$ and $z_{\mathrm{max}}$, and 100 $k$ values between $k_{\mathrm{min}}$ (which is survey dependent) and $0.075~h/\mathrm{Mpc}$.
We verified that doubling the number of $z$ or $\mu$ bins or $k$ points beyond these values changes the final forecasted error for the single-tracer forecasts by less than 1\%.
To be consistent with Ref.~\cite{Noah21}, we perform all single-tracer forecasts with the UMF prediction with $p=1$.
For a more detailed discussion of a careful treatment of the choice of $p$ in the context of real data, including the choice of priors, we refer the reader to Ref.~\cite{2022JCAP...11..013B}.
We neglect wide-angle effects and large-scale relativistic effects on the galaxy power spectrum.
We neglect any additional systematic effects that could contaminate a large-scale measurement of the power spectrum (e.g. stellar contamination), though it is of course extremely important to characterize these well in a real data analysis.

Here we describe the galaxy samples used in our forecasts.
The primary samples we consider are the Dark Energy Spectroscopic Instrument (DESI) Luminous Red Galaxies (LRGs) and Emission Line Galaxies (ELGs).
For the DESI LRG-like sample we use the redshift distribution of \cite{rongpu_lrg}, and for the DESI ELG-like sample we use the redshift distribution of \cite{AnandELG}.
For DESI $f_{\mathrm{sky}}= 0.34$ \cite{desi2016doc}, $k_{\mathrm{min}} = 0.0023~h/\mathrm{Mpc}$, and $z_{\rm{min}} = 0.0$, $z_{\rm{max}} = 2.0$.
For the linear galaxy bias of the two DESI samples in the single-tracer analysis, we make the empirically calibrated choice \cite{desi2016doc,dzDESIbiasrelations} of $b_{\rm{LRG}} = \frac{1.7}{D(z)}$, $b_{\rm{ELG}} = \frac{0.84}{D(z)}$, where $D(z)$ is the linear growth factor (normalized to $1$ today) computed in our fiducial cosmology.
For the multi-tracer forecasts using the results of Section.~\ref{sec:GNN}, we assume the results evolve similarly with $D(z)$ (i.e. $b(z) = \frac{D(z=1)}{D(z)}b$, and similarly for $\bphi$).

For the MegaMapper forecasts, we follow the prescriptions of Refs.~\cite{CabassMegaMapper,Noah21,simone_highz_spectro,mike_martin_megamapper} for the galaxy density $n(z)$ and linear bias $b(z)$, we assume $f_{\mathrm{sky}}= 0.34$, $k_{\mathrm{min}} = 0.0017~h/\mathrm{Mpc}$, and $z_{\rm{min}} = 2.0$, $z_{\rm{max}} = 5.0$.
For SPHEREx, we use $f_{\mathrm{sky}}=0.65$, $k_{\mathrm{min}} = 0.001~h/\mathrm{Mpc}$, and $z_{\rm{min}} = 0.1$, $z_{\rm{max}} = 3.0$.
For the SPHEREx forecasts, we assign linear bias as a function of redshift based on the procedure outlined in Ref.~\cite{spherex14}, but use the fitting functions of Ref.~\cite{bhattacharya11} as implemented in \texttt{COLOSSUS} \cite{colossus} (and have verified that this choice changes the halo bias of individual redshift error samples by less than 1\% with respect to the fitting functions of Refs.~\cite{tinker08,tinker10}).
We consider the redshift-error samples (labeled by $\sigz$) on their own and in combination.
We incorporate the SPHEREx redshift errors as described in Ref.~\cite{spherex14} for the various redshift error samples, multiplying the shot noise term by $\exp(-\left[k \mu \sigma_{z} \frac{d\chi}{dz}\right]^{2})$.
All non-SPHEREx forecasts in this paper assume zero redshift error.
When combining several redshift error samples, we always use the largest redshift error.
For the fully-combined SPHEREx sample, this corresponds to an assumed redshift error of $\sigz = 0.2$, so we deviate slightly from the treatment of Ref.~\cite{Noah21}, who used $\sigz = 0.05$ for this full sample, and therefore we are more pessimistic in our forecasts.
We compute an effective bias for combined samples by weighting the bias by the sample number density.

We work in linear theory and at a fiducial cosmology where $\fnlloc=0$, and so use the Fisher matrix for $\delta_{X}(k,\mu,z)$ which, in the Gaussian approximation is \cite{1997ApJ...480...22T}:
\begin{equation}
    F_{\alpha\beta} = \sum_{i_{z} = 1}^{N_{z}-1} \int_{0}^{1} d\mu \int_{k_{\mathrm{min}}}^{k_{\mathrm{max}}} N_{k,i_{z}} \mathrm{Tr}\left[\mathbf{C}_{,\alpha}\mathbf{C}^{-1} \mathbf{C}_{,\beta} \mathbf{C}^{-1}\right]
    \label{eqn:Fisher} 
\end{equation}
where $\mathbf{C}$ can be either the two-tracer version $\mathbf{C}^{(2)}$ or the three-tracer version $\mathbf{C}^{(3)}$, with
\begin{equation}
\mathbf{C}^{(2)} = 
\begin{pmatrix}
C_{XX} & C_{XY} \\
 & C_{YY} 
\end{pmatrix},\quad
\mathbf{C}^{(3)} = 
\begin{pmatrix}
C_{XX} & C_{XY} & C_{XZ} \\
 & C_{YY} & C_{YZ} \\
 & & C_{ZZ}
\end{pmatrix},
\label{eqn:covariance}
\end{equation}
where the field covariance and number of modes $N_{k,i_{z}}$ in a volume shell $V_{i_{z}}$ are
\begin{align}
C_{XY} &= \left(b^{X} + \mu^{2} f + \bphi^{X}\fnlloc \mathcal{M}^{-1}(k) \right)\left(b^{Y} + \mu^{2} f + \bphi^{Y}\fnlloc \mathcal{M}^{-1}(k) \right) P_{L}(k) + \frac{\delta^{D}_{XY} }{\bar{n}_{X} } 
,\\
N_{k,i_{z}} &= \frac{dk k^{2} V_{i_{z}} }{2\pi^2},
\end{align}
respectively, and we neglect both tracer non-Poisson stochasticity and cross-stochasticity between different tracer fields \cite{baldauf_stoch,hamaus_halo_forecast,ginzburg_desjacques_multi_sn}.
Here $i_{z}$ gives the index of the $z-$bin midpoint used to compute Fisher matrix element, and $V_{i_{z}}$ is a spherical shell computed using the adjacent redshift bin edges and is scaled by $f_{\mathrm{sky}}$.

Here, in the notation of Ref.~\cite{UrosMulti}, $\alpha = \frac{b_{1}}{b_{2}}$ and $P_{2} = b_{2}^{2} P$ (at our fiducial value of $\fnlloc=0$), where $b_{1}$ and $b_{2}$ are the \textit{linear} biases of the first and second samples under consideration.
We define $\beta \equiv \frac{b_{3}}{b_{2}}$ analogously for the three-tracer case.
Explicit expressions for the Fisher matrix element $F_{\fnlloc \fnlloc}$ are provided in Appendix~\ref{app:fisher_triple} and in the public code accompanying this paper (linked above).

\subsection{Results for $\sigma(\fnlloc)$}
\label{subsec:fisher_results}

\subsubsection{DESI}
\begin{table}[h!]
    \centering
    \begin{tabular}{|c|c|c|c|c|c|}
    \hline
    Sample & $\sigma_{\rm{ST}}(\fnlloc)$ &  $\sigma_{2,\rm{MT}}(\fnlloc)$ &  $\sigma^{(P)}_{2,\rm{MT}}(\fnlloc)$ &  $\sigma_{3,\rm{MT}}(\fnlloc)$ &  $\sigma^{(P)}_{3,\rm{MT}}(\fnlloc)$ \\
    \hline
    ELG, ideal & 7.1 & 1.3 & 1.4 & 1.7 & 1.5\\
    ELG, pred. & . & 10 & 2.5  & 12 &  2.3\\
    LRG, ideal & 5.0 & 3.0 & 2.6 & 3.5 & 3.4\\
    LRG, pred. & . & 16 & 3.7 & 17 & 4.7\\
    \hline
    \end{tabular}
    \caption{Local primordial non-Gaussianity forecasts for single- and multi-tracer forecasts for ELG and LRG-like simulated galaxies. For each column $\sigma_{i,XT}^{(P)}$, $i$ denotes the number of tracers used in each forecast, $(P)$ indicates whether the forecast includes $P_{2}$ as a parameter, and $X=\{S,M\}$ is $S$ for single-tracer and $M$ for multi-tracer. For the 2-tracer forecasts, we use the highest and lowest $\bphi$ tertiles.}
    \label{tab:multi_choices}
\end{table}

Here we present $\sigma(\fnlloc)$ forecast results - we first perform forecasts for LRGs and ELGs individually with several multi-tracer setup choices (Table~\ref{tab:multi_choices}), then present results illustrating the effect of selection choice on $\sigma(\fnlloc)$ for ELGs (Table~\ref{tab:multi_num_ssfr}) before finally considering the case of ELGs and LRGs in their overlap region (Table~\ref{tab:elg_x_lrg}).

Table~\ref{tab:multi_choices} shows the improvement of the multi-tracer forecasts over the single-tracer forecasts for several choices of multi-tracer setup for our fiducial choice of DESI mock galaxy selection ($\bar{n}_{\mathrm{ELG}} = 5 \times 10^{-4}~[h^{-1}~\mathrm{Mpc}]^{-3}, \log_{10}\left(\frac{sSFR}{h~yr^{-1}}\right)=-9.09$).
Here we report results for both the simulated LRG and ELG galaxy samples, for the ideal values of $\bphi(b)$ in each tertile (i.e. concentration information is recovered perfectly) and for the predicted (``pred.'') values using the machine learned relationship $\bphi(b)$ from observable galaxy properties.
While clearly the ML model cannot access all the $\fnlloc$ information contained in halo mass and concentration, it recovers enough information to significantly improve the forecasted LPNG amplitude error $\sigma(\fnlloc)$.
The improvement for the predicted relations is largest for ELGs, where we can obtain a forecasted error of $\sigma(\fnlloc) = 2.3$, a factor of 3 improvement over the single-tracer case.

We show both 2-tracer and 3-tracer forecasts.
The 2-tracer forecasts use the upper and lower tertiles of the three-bin splits, while the 3-tracer forecasts use all three tertiles (both cases use number density $\bar{n}/3$ for each tracer)
When making multi-tracer forecasts, we compute the error on $\fnlloc$ using two different procedures - first, when the only parameters considered in the Fisher matrix is (are) the relative amplitude(s) $\alpha, ~(\beta)$, and second, when both the relative amplitude(s) $\alpha, ~(\beta,)$ and $P_{2}$, the power spectrum corresponding to the second sample are used.
Since in the first case we use only the relative amplitude for the two-tracer forecasts, we refer to the case of including $P_{2}$ as a parameter as ``$(P)$'', since we are including the power spectrum as a parameter.
Including $P_{2}$ as a parameter generally reduces the forecasted error, however, in the sample-variance limit $P_{2}$ adds no information (see discussion around eqn.~\ref{eqn:fish_info_23_rat_1}).
In multi-tracer forecasts used in the rest of this paper, we include $P_{2}$ as a parameter.
Some brief further discussion of these forecasting aspects can be found in Appendix~\ref{app:fisher_triple}.

\begin{table}[h!]
    \centering
    \begin{tabularx}{\linewidth}{|c | X |}
    \hline
    $\log_{10}\left(\frac{sSFR}{h~\mathrm{yr}^{-1}}\right)$ & \hfill -9.09~ \hfill   -9.23  \hfill\null \\
    \hline
    $n_{\mathrm{ELG}}~[h^{-1}~\mathrm{Mpc}]^{-3}$  & 
    \hfill ideal \hfill pred. \hfill ideal \hfill pred. \hfill\null \\
    \hline
        \makecell{$5\times 10^{-4}$\\ $7\times 10^{-4}$ \\ $1\times 10^{-3}$} & \hfill 
 \makecell{1.5 \\ 1.6 \\ 1.6} \hfill \makecell{2.3 \\ 2.3 \\ 2.5} \hfill \makecell{1.4 \\ 1.5 \\ 1.5} \hfill\makecell{2.3 \\ 2.2 \\ 2.2} \hfill\null \\
    \hline
    \end{tabularx}
    \caption{Multi-tracer $\sigma(\fnlloc)$ Fisher forecasts for several choices of $\bar{n}_{g}$ and sSFR for DESI ELG-like simulated TNG galaxies. 
    Results do not vary significantly with the choice of mock galaxy selection.}
    \label{tab:multi_num_ssfr}
\end{table}

Table~\ref{tab:multi_num_ssfr} characterizes the effect of the simulated galaxy sample selection strategy on the forecasts for $\sigma(\fnlloc)$.
These selections were mentioned already in Section~\ref{sec:GNN} (in the context of the learned $\bphi(b)$ relation), but here we supplement this by checking the effect of the choice of specific star formation rate split and stellar mass threshold, which determines the number density for a fixed sSFR split, on the final forecasted $\fnlloc$ error.
We find that the final results are mostly insensitive to these choices, as they vary only by a factor of 17\% or less (within the ideal or predicted cases).
It may appear counter-intuitive that $\sigma(\fnlloc)$ increases even in the ideal case for both specific star-formation rates as the number density increases, but this is due to the fact that the mean linear bias and, at first approximation, the LPNG bias both shift downward as the number density increases.
In more detail, while the $\bphi$ values drop uniformly across all tertiles with the mean, the $b$ value in the highest tertile hardly changes with increasing number density, but the lower tertile drops significantly, and this leads to a lower change in linear bias between the upper and lower tertiles at higher number density, which in turn leads to a smaller relative amplitude Jacobian factor in the Fisher matrix (see eqn.~\ref{eqn:jac_alpha}).
In any event, the highest difference of $17\%$ in $\sigma(\fnlloc)$ between the two choices of sSFR is for the highest number density sample (which is not used elsewhere).

\begin{table}[h!]
    \centering
    \begin{tabular}{|c|c|}
    \hline
    ELG + LRG & $\sigma(\fnlloc)$\\
    \hline
        $p=1$ & 4.0\\
        (2) ($\overline{\mathrm{LRG}}$, $\overline{\mathrm{ELG}}$), ideal & 2.3\\
        (2) ($\overline{\mathrm{LRG}}$, $\overline{\mathrm{ELG}}$), pred & 2.3\\
        (2) (LRG+, ELG+), ideal & 1.4 \\
        (2) (LRG+, ELG+), pred & 2.4 \\
        (3) (LRG-, LRG+, ELG-), ideal & 0.8 \\
        (3) (LRG-, LRG+, ELG-), pred & 2.0 \\ 
        (3) (LRG-, ELG+, ELG-), ideal & 0.8 \\
        (3) (LRG-, ELG+, ELG-), pred & 2.0 \\
        (3) (LRG-, ELG+, else), ideal & 0.6 \\
        (3) (LRG-, ELG+, else), pred & 1.5 \\
    \hline
    \end{tabular}
    \caption{Local primordial non-Gaussianity amplitude $\fnlloc$ multi-tracer forecasts for ELG-LRG in their overlap region.
    The entries listed ``($\overline{\mathrm{LRG}}$, $\overline{\mathrm{ELG}}$)'' use the mean $\bphi(b)$ prediction with the full number density of each sample
    Here ``+'' and ``-'' denote the top and bottom tertile values of $\bphi(b)$ of the respective samples, respectively. 
    The row entries following the ($\overline{\mathrm{LRG}}$, $\overline{\mathrm{ELG}}$) rows use ($i$)-tracer forecasts ($i=2,3$) with number densities of $\bar{n}/3$ for each sample.
    In the final two rows, the unused subsamples are combined with bias weighted by their number densities.
    }
    \label{tab:elg_x_lrg}
\end{table}

Table~\ref{tab:elg_x_lrg} shows $\sigma(\fnlloc)$ Fisher forecasts for several multi-tracer sample configurations drawn from the ELG and LRG simulated galaxy samples.
To represent the simplest possible multi-tracer forecast setup using the undivided LRG and ELG samples, we report $\sigma(\fnlloc) = 4.0$ for the UMF bias prediction (eqn.~\ref{eqn:umf}) with the parameter $p$ set to $1$.
This forecast does not make use of the learned $\bphi(b)$ relations presented in this paper.
We next consider a similar case that uses the mean of the $\bphi(b)$ values for the simulated LRG sample and the simulated ELG sample (with the fiducial choice of sample parameters: $\bar{n}_{ELG} = 5 \times 10^{-4}$, $\bar{n}_{LRG} = 2 \times 10^{-4}$, and $\log_{10}(\mathrm{sSFR}) = -9.09$).

When we consider the multi-tracer forecasts that use two samples split by $\bphi$ tertile, we find a lower $\sigma(\fnlloc)$.
When using the upper tertile samples for both ELGs and LRGs (LRG+,ELG+), we find a factor of 2.9 reduction in $\sigma(\fnlloc)$ compared to the $p=1$ UMF forecast in the ideal case, where concentration and halo mass information are perfectly recovered, and a factor of 1.7 reduction in $\sigma(\fnlloc)$ when using the learned relation.
We also consider three-tracer forecasts for the ELG and LRG tertile samples.
Using the lower and upper tertiles for the LRGs and lower tertiles for the ELGs (LRG-,LRG+,ELG-), we find smaller forecasted errors than in the two-tracer case, and using the lower and upper tertiles for the ELGs along with the lower tertile for the LRGs (LRG-,ELG+,ELG-) provides similar results, reducing the forecasted error by a factor of 5 compared to the $p=1$ UMF forecast in the ideal case and a factor of 2 in the predicted case.

Finally, based on the promising results of the (LRG-,ELG+,ELG-) choice of subsamples, we show in the last two rows of Table~\ref{tab:elg_x_lrg} forecasts for the case where ELG- in this triplet is replaced with all the other sub-samples combined.
In particular, the third sample is the combination of ELG-, ELG0, LRG0, and LRG+, where ``0'' indicates the central tertile.
Bias values for these samples are weighted by their number density (as for the SPHEREx redshift error $\sigz$ samples).
In this case, we find the most constraining forecasts we report for LRGs and ELGs, with $\sigma(\fnlloc) = 0.6$ and $\sigma(\fnlloc) = 1.5$ for the ideal and predicted $\bphi$ values, which are now factors of 6.7 and 2.7 reductions over the naive $p=1$ multi-tracer forecast, respectively.

We do not list all possible permutations of the ELG and LRG tertile samples (and their combinations) for brevity, though some of the other combinations produce a lower $\sigma(\fnlloc)$ in the predicted case than we show here.
It would also be interesting to further consider optimally combining split subsamples in general \citep[e.g. as for mass in][]{slosar_optimal,hamaus_halo_forecast} - we took a first step in exploring this here by using the ``else'' samples in Table~\ref{tab:elg_x_lrg}, which produces excellent results.

We now discuss the origin of the sizeable reduction in our forecasts for $\sigma(\fnlloc)$ over the single-tracer case.
The Jacobian factor involved in converting from the Fisher information on the relative amplitude multi-tracer parameters ($F_{ij}$ where $i,j \in \{\alpha, \beta \}$) to that of LPNG ($F_{\fnlloc \fnlloc}$) involves the factor (reproduced from Appendix~\ref{app:fisher_triple}):
\begin{equation}
    \frac{\partial \alpha}{\partial \fnlloc} = \alpha \left(\frac{b_{\phi,1}}{b_1} - \frac{b_{\phi,2}}{b_2} \right) \mathcal{M}^{-1}(k),
\label{eqn:jac_alpha}
\end{equation} 
which is evaluated at $\fnlloc=0$.
For a fiducial choice of $\fnlloc=0$, the Jacobian factor of eqn.~\ref{eqn:jac_alpha} is in fact the only place that the LPNG bias enters the Fisher forecast calculation.
A similar expression holds for $b_3,b_{\phi,3}$ and $\beta$ in the 3-tracer case.

If we consider the limit where the relative amplitude $\alpha\to1$, then clearly this factor is proportional to $\Delta \bphi$.
This limit is exact in a 2-way split of a single halo mass bin if linear bias does not evolve across the bin or with the splitting parameter.

We now make the even further simplification, for illustrative purposes, that each $\bphi$ of the two samples in question is described by a UMF relation for some choice of the parameter $p$, and so $b_{\phi,1} = 2\delta_{c}(b_{1}-p_{1})$, $b_{\phi,2} = 2\delta_{c}(b_{2}-p_{2})$, and for $\alpha\to1$ we have $\Delta\bphi \to 2\delta_{c}\Delta p$, and $\frac{\partial \alpha}{\partial \fnlloc} \to -2\delta_{c} \Delta p \mathcal{M}^{-1}(k)$.
From this demonstration, it would appear that we can drive our forecasted error on LPNG as low as we wish as we take $\Delta p \to \infty$.
We further discuss the implications of this apparent conclusion in the next section.

\subsubsection{Future LPNG surveys}
\begin{figure}[h!]
    \centering
    \includegraphics[width=0.8\textwidth]{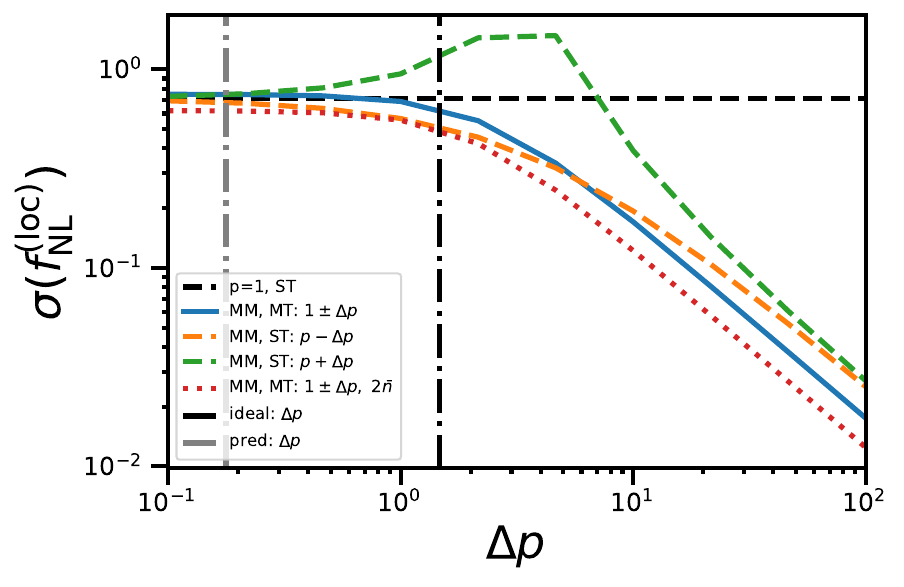} 
    \caption{(\textit{MegaMapper}): The dependence of forecasted $\sigma(\fnlloc)$ multi-tracer forecasts on $\Delta p$, the difference in the UMF parameter, for the planned MegaMapper survey. 
    The black dashed line shows the single-tracer (ST) forecast with the original UMF scenario, for which $p=1$.
    The solid blue line shows the multi-tracer (MT) forecast for the case in which the full MegaMapper sample is split into tertiles and the top and bottom tertiles are used with equal number density bins of $\frac{\bar{n}}{3}$, each of which is assigned $\bphi = 1 \pm \Delta p$.
    For comparison, the dashed colored lines show the single tracer forecasts with $p=1+\Delta p$ (green dashed) and $p=1-\Delta p$ (orange dashed).
    The red dotted line is the same as the blue line (multi-tracer forecast), but with twice the number density.
    The horizontal dash-dotted lines in gray and black show the approximate values of $\Delta p$ for the learned and ideal $\bphi$ values for the mock DESI galaxy samples.
    }
    \label{fig:future_deltap_mm}
\end{figure}

To explore the potential of using samples with large differences in $\bphi$ further, we apply our multi-tracer forecasting framework to two future surveys that will target LPNG, MegaMapper \cite{simone_highz_spectro}, and SPHEREx \cite{spherex14} using the setup described in Section~\ref{subsec:fisher_setup}. 
We use the UMF relations as input for $\bphi(b_{1})$ in these forecasts,
as we do not have the information necessary to construct detailed simulated galaxies and corresponding learned $\bphi(b_{1})$ for the future MegaMapper or SPHEREx galaxy populations.
Since we do not have a solid basis for selecting $\Delta p$ for these forecasts, we instead consider forecasts at a range of $\Delta p$.
For the following multi-tracer forecasts, we will set $b_{\phi,i} = 2\delta_{c}(b_{i} -[p \pm \Delta p])$ where $p=1$ unless otherwise indicated\footnote{This is in a sense a somewhat pessimistic choice, see Appendix~\ref{app:fisher_triple} for more discussion}.
The value $p=1$ is the best choice to make contact with the literature, which mostly uses the UMF with $p=1$.
In principle, if we were to choose a lower $p$, we should only reduce $\sigma(\fnlloc)$ (see Fig.~\ref{fig:future_deltap_mm})  - this is closely related to the zero-bias effect discussed by Ref.~\cite{ZeroBias}.

\begin{figure}[h!]
    \centering
    \includegraphics[width=0.8\textwidth]{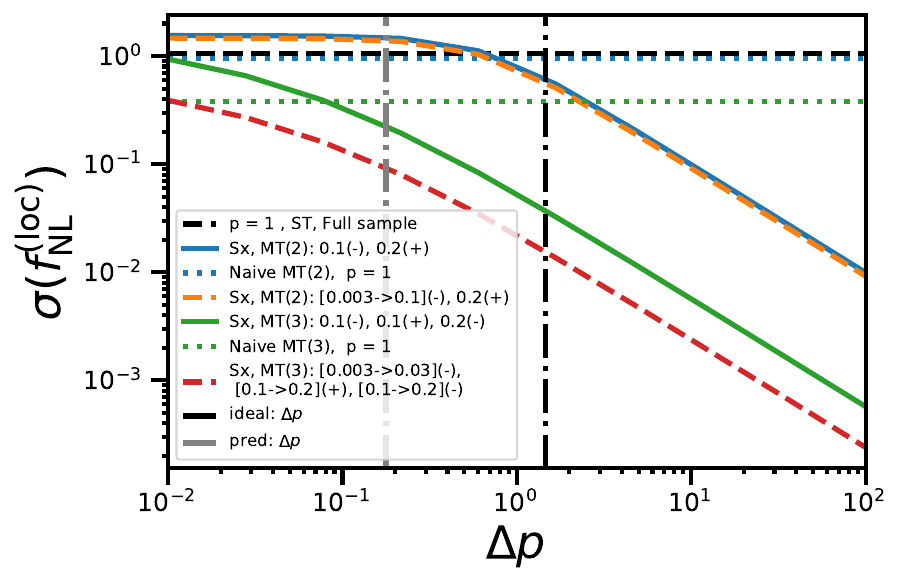} 
    \caption{(\textit{SPHEREx}): Similar to Figure~\ref{fig:future_deltap_mm}, but for multi-tracer forecasts involving multiple SPHEREx redshift error samples.
    The black dashed line again shows the UMF single-tracer forecast.
    The solid blue line shows the two-tracer (2) multi-tracer (MT) forecast for the $\sigz = 0.01$ and $\sigz = 0.2$ redshift error samples, where the former has UMF parameter $p=1-\Delta p$ (``-'') and the latter has $p=1+\Delta p$ (``+'').
    The orange dashed line shows the two-tracer MT forecast where the first tracer is the lowest redshift error sample ($\sigz = 0.003$) with $p=1-\Delta p$ and the second tracer is the combined sample of all other redshift error samples (with $p = 1 + \Delta p$).
    The green solid line shows the three-tracer (3) multi-tracer forecast for the three splits of the redshift error samples - the $\sigz = 0.01$ sample with $p = 1-\Delta p$ (-), the $\sigz= 0.1$ sample with $p = 1+\Delta p$ (+), and the $\sigz = 0.2$ sample with $p = 1-\Delta p$ (-).
    The red dashed line shows another three-tracer (3) forecast using combined samples - the combined $\sigz = 0.003, 0.01$ samples with $p = 1-\Delta p$ (-), the combined $\sigz = 0.03, 0.1, 0.2$ samples with $p = 1+\Delta p$ (+), and the second half of the same sample with $p = 1-\Delta p$ (-).
    Again, horizontal dash-dotted lines in gray and black show the approximate values of $\Delta p$ for the learned $\bphi$ for the mock DESI galaxy samples.
    }
    \label{fig:future_deltap_sx}
\end{figure}

Figure~\ref{fig:future_deltap_mm} shows the resulting dependence of the two-tracer forecasted error $\sigma(\fnlloc)$ on $\Delta p$ for MegaMapper.
The solid blue line shows the two-tracer forecast for the scenario in which the number density of the MegaMapper sample is split into thirds, and the top and bottom tertiles of the sample are assigned the \textit{same} linear bias $b$ and an LPNG bias of either $\bphi = 2\delta_c (b -(1 \pm \Delta p))$.
When $\Delta p \to 0$, we find that we approach the single-tracer results at $p=1$, (since for $\Delta p =0$ the relative amplitude Jacobian factors vanish and we are left with only $P_{2}$ information). 
It is immediately apparent that there is no bound on how much the multi-tracer forecast can improve upon the single-tracer forecast at fixed $p$.

To make a fairer comparison to the single tracer case, we also plot in the dashed lines the single tracer forecasts with $p=1+\Delta p$ (green) and $p = 1-\Delta p$ (orange).
We can see that by drastically increasing or decreasing $p$ in the single-tracer case we can obtain a large reduction in $\sigma(\fnlloc)$ that is qualitatively similar to that of the multi-tracer case.
In the limit of $\Delta p \to \infty$ we see that the multi-tracer case always outperforms the single-tracer case by a roughly constant factor, and both lines have similar slope.
This is expected due to SV cancellation and the fact that (for fixed $b$) in the large-$p$ limit, we have that $\sigma_{ST}(\fnlloc) \propto \frac{b}{p}$, and for large $\Delta p$, $\sigma_{MT}^{(2)}(\fnlloc) \propto  \frac{b}{\Delta p}$.
These single-tracer forecasts use the full number density of the sample, so to account for this difference when comparing to multi-tracer, we also artificially increase the number density by a factor of two in the multi-tracer case as an additional point of comparison (red dotted line).
Here we can see that the slight improvement of single-tracer over multi-tracer around $\Delta p =1$ is due to the reduced number density of the multi-tracer sample.

We also mark in vertical dash-dotted gray and dash-dotted black the values of $\Delta p$ that are close to describing $\bphi$ for the upper and lower tertiles of the ELG and LRG simulated galaxies in the predicted and ideal cases, respectively\footnote{These values are simply calculated as $\Delta p = \frac12 (\delta^{c})^{-1}(b_{\phi}^{+}-b_{\phi}^{-})$, so there is a symmetry here that isn't present in the ELG/LRG results. For more discussion see Appendix~\ref{app:fisher_triple}}.
While this exercise illustrates in part where the improved constraints for DESI galaxies is coming from, the change in $\Delta p$ is of course not the only thing driving a difference in the mulit-tracer and single-tracer forecasts.
As in the usual halo-mass-based multi-tracer application, $b_{1}$ still plays a significant role in both the Jacobian factor of eqn.~\ref{eqn:jac_alpha} and the Fisher matrix elements of the multi-tracer parameters (e.g. through factors of the form $b^2 P \bar{n}$).
For anything but the idealized sample-variance limit, the increased number density from splitting the full sample can also strongly affect the final forecasted error.
Further study of physical and practical limits on $\Delta p$ will be the subject of future work.

We performed a similar exercise for SPHEREx in Figure~\ref{fig:future_deltap_sx}, which shows forecasted $\sigma(\fnlloc)$ as a function of $\Delta p$ for several combinations of SPHEREx redshift error samples.
In black dashed we again show the single-tracer forecast for $p=1$ for the combined sample of all SPHEREx redshift error samples (using the most pessimistic redshift error).
Multi-tracer forecasts for individual redshift error samples (labeled by $\sigz \in \{0.003,0.01,0.03,0.1,0.2\}$) where we assign splits on $\Delta p$ are shown by the solid lines.
We also show ``naive'' two-tracer forecasts when considering different redshift error samples using the same $p$ (but these samples have different $b$). 
To take advantage of the high number density of SPHEREx, we also show multi-tracer forecasts for combinations of redshift error samples (for example, combining the two lowest redshift error samples to create one multi-tracer sample and combining the three highest redshift error samples to create the second multi-tracer sample) as the dashed lines.
Here the ``MT($i$)'' and $(\pm)$ labels serve the same purpose as in Table~\ref{tab:elg_x_lrg}, indicating how many tracers are used in the forecast and whether the top or bottom tertile of the $\Delta p$-split sample is used. 

We find similar results to Fig.~\ref{fig:future_deltap_mm}, in that the multi-tracer forecasts improve over the single-tracer forecasts as $\Delta p$ increases.
First, we discuss the case where we use individual redshift error samples as the multi-tracer samples (solid lines).
The naive forecasts (dotted lines) are also for the individual redshift samples, and should be compared to lines of the same color.

At low $\Delta p$ the improvement of the two-tracer $\Delta p$ multi-tracer forecast over the full single-tracer forecast (black-dashed) with the full sample disappears since both are dominated by the most pessimistic redshift error of $\sigz =0.2$. 
This is in fact worse than the naive multi-tracer case (dotted blue line) due to the lower number density.
Similar statements hold for the three-tracer case, though the three-tracer results show what is effectively only the high $\Delta p$ regime of the 2-tracer case, and, we can see that the difference between the naive and $\Delta p$-split three-tracer forecasts is larger than in the two-tracer case.

Turning now to the combined samples, we see that combining several redshift samples (with the appropriated weighted linear biases) leads to a reduction in shot noise\footnote{The redshift error is unaffected, as we always choose the most pessimistic redshift error for a combined sample.}  due to increased number density of the sample, and this gives lower $\sigma(\fnlloc)$ when $\Delta p \to 0$ as seen in the difference between solid and dashed multi-tracer lines.
We can also see from the red dashed line that the improvement over single-tracer is similar to the red dotted curve in Fig.~\ref{fig:future_deltap_mm}, again suggesting that the uniform reduction in $\sigma(\fnlloc)$ over the three-tracer single sample forecast (green solid line) is due to the increased number density of the combined samples.
Essentially all of the lowest error multi-tracer forecasts come from the high redshift error samples (or combinations that use the highest redshift error), as these have the highest number density (roughly $2-3 \times 10^{-3}~[h^{-1}~\mathrm{Mpc}]^{-3}$), largest redshift range (extend out to $z=3$), and the measurement of $\fnlloc$ is relatively insensitive to these errors.

\subsection{Discussion} 
\label{subsec:discussion}

We are leveraging information about $\bphi$ through the portal of halo mass and concentration, but in galaxy survey data we only have access to observable galaxy properties, not the concentration itself. 
We have fixed the galaxy-halo connection (GHC) that maps the halo field to the galaxy field for a specific galaxy sample to that afforded by the IllustrisTNG galaxy formation model.
Even assuming this fixed GHC, we will inevitably lose information about the halo mass and concentration when passing to simulated galaxies.
For this reason, we report multi-tracer forecasts for both ``ideal'' and ``predicted'' $\bphi$ for the ELG and LRG galaxy samples, where the predicted case captures the limitations of using only observable galaxy properties rather than host halo properties.
The former refers to the (fictitious) scenario in which we could perfectly recover $\bphi$ from the simulated galaxy sample based on mass and concentration, and the latter to what the ML model of Section~\ref{sec:GNN} recovers.
The ``ideal'' forecasts results should therefore be interpreted as quantifying the maximal $\bphi$ information available from the given galaxy sample, while the ``predicted'' forecasts tell us what information about $\bphi$ can be obtained by applying the ML predictions directly to DESI data (assuming the TNG galaxy formation model, and that the observable input features have actually been measured).
Following this initial work, several strategies could be employed to improve the ML prediction of $\bphi$. 
The larger MilleniumTNG simulation \cite{pakmor_milleniumTNG} could be used, which would increase the size of the datasets used by a factor of almost 15.  
We also found, as expected, that results were much improved in real space compared to redshift space, so adding a reconstruction procedure \cite{eisenstein_recon} could potentially improve results.
Another option would be to use additional galaxy properties that are more challenging to model as input features, such as stellar half-mass radius, metallicity, or maximum value of spherically-averaged rotation curve.

We have made several simplifying assumptions in this work when applying the $\bphi(b_{1})$ results of Section~\ref{sec:GNN} to the Fisher forecasts of this section.
We have ignored the impact of fiber collisions, which, for real galaxies, would impact the ability to determine close pairs of galaxies and to compute the local environment statistics used for determining $\bphi$ from observables.
However, we expect that since local environment features are significantly less informative than the magnitude bands and stellar mass, including fiber collisions would have a minimal impact on our results. 
We used the concentration-LPNG bias relation of Ref.~\cite{Lazeyras22} which uses different simulations than those of IllustrisTNG itself, but since Ref.~\cite{Lazeyras22}  uses logarithmized concentration, we do not expect a significant effect from this choice.
We neglected the exact redshift dependence of the linear bias $b$, instead using the linear growth factor to account for the redshift dependence as in Ref.~\cite{desi2016doc}.
While here we only worked with halo concentration and mass, there is certainly no a priori reason to ignore halo properties such as age (e.g. vis a vis recent mergers or formation time \cite{Slosar08,reid10_assmblyhistory}), and future work should further address the connection of these properties with $\fnlloc$.
The most significant limitation of this work is that we used only the IllustrisTNG galaxy formation model - further work should of course be done to consider alternative galaxy formation models to establish the robustness of (or lack thereof) the ability to calibrate $\bphi(b)$ using observable galaxy properties.
By using IllustrisTNG galaxy formation model only and working with halo $\bphi$, we also neglect the response of galaxy halo occupation to LPNG, which has been shown to be non-negligible \cite{2020JCAP...12..013B,marinucci_sham,voivodic_hod}.

While we were finishing this manuscript, the preprint of Ref.~\cite{barreira23} appeared presenting a two-tracer investigation of two IllustrisTNG galaxy populations selected by various secondary properties (such as halo concentration and $g-r$ color).
While the focus of our work and Ref.~\cite{barreira23} are similar, here we focus on DESI ELG and LRG galaxies in redshift space, specific future survey forecasts, and report $\sigma(\fnlloc)$, rather than the improvement over single-tracer information.
We consider both Ref.~\cite{barreira23} and our work to be complementary, as Ref.~\cite{barreira23} provides a clear and simplified explanation of the results of Ref.~\cite{UrosMulti} applied to secondary galaxy properties, while also characterizing systematic effects and priors on $\bphi$, which we largely ignore here. 
In the other direction, we have provided a significantly more realistic treatment of how splitting by $\bphi$ on a secondary parameter (in the special case of halo concentration) can be used in a multi-tracer setting for current and upcoming spectroscopic surveys.

We also developed the machine learning tools described in Section~\ref{sec:GNN}, which incorporate several observable quantities to predict $\bphi$.
Our ability to constrain $\fnlloc$ is, however, limited by the dependence of $\bphi$ on concentration and mass, while Ref.~\cite{barreira23} makes use of the dependence $\bphi$ on quantities such as $(g-r)$ color directly through hydrodynamical Separate Universe simulations.
It would be interesting to go beyond both our work here and the work of Ref.~\cite{barreira23} to further explore the optimal choice of observable characteristics by which to split a tracer population to obtain maximally different $\bphi$ values with ML methods using different galaxy formation models.

\section{Conclusions}
\label{sec:conclusions}

Galaxy surveys contain information about the field content of the inflationary universe through the influence of local Primordial non-Gaussianity on dark matter halo formation.
To extract this information, we must assume a relationship between the number density of LSS tracers and primordial fields, which is quantified by the LPNG bias parameter $\bphi$.
In this work, we trained a machine learning model to connect this parameter to observable properties of DESI galaxies by way of host halo mass and concentration using simulated IllustrisTNG galaxies.
We then used the learned relationship between the linear bias $b_{1}$ and LPNG bias $\bphi$ for simulated galaxies to perform multi-tracer Fisher forecasts for the amplitude of LPNG $\fnlloc$ in DESI using galaxy power spectra.
We also illustrated that for future LSS galaxy surveys that will target LPNG (MegaMapper, SPHEREx), there is potentially significant untapped information on $\fnlloc$ that can be extracted with the multi-tracer method when the samples can be split by host halo concentration.

\noindent
We summarize our major conclusions here:
\begin{itemize}
    \item We find that when considering simulated IllustrisTNG galaxies in a realistic setting (with observed DESI number densities, and including redshift-space distortions), the ML method can only extract part of the information about halo concentration from observable properties. This illustrates the challenge of using (unobservable) halo properties for accessing information on LPNG. We may however be limited by the small 
    size of IllustrisTNG training data, 
    rather than by the 
    predictive power of observables, in which case our predictions 
    are conservative. 
    \item Nevertheless, our ML method extracts predictions for $\bphi$ that contain enough information to allow multi-tracer Fisher forecasts of $\sigma(\fnlloc)$ to greatly improve upon single-tracer forecasts. For certain sub-sample selections of DESI ELGs and LRGs, we find reductions of $\sigma(\fnlloc)$ that are frequently a factor of 50\%, and up to a factor of 3.
    \item We also provide forecasts for the ``ideal'' case in which $\bphi$ information is perfectly known for each halo.
    In that case, even further reductions in $\sigma(\fnlloc)$ are possible - up to a factor of 5, motivating further improvement of the methods developed here.
    \item We argue that future spectroscopic surveys targeting PNG (such as SPHEREx and MegaMapper) could greatly improve their constraining power on LPNG by employing a strategy similar to the one presented here, and support this argument with UMF-based forecasts for these surveys using varying $\Delta p$. More careful investigation of the potential sub-sample multi-tracer gains in LPNG information from these surveys is required.
\end{itemize}

It will be interesting to further explore to what extent the forecasts we have presented are attainable in real data analyses, and to what extent it is physically possible to obtain very large values of $\Delta \bphi$ in cosmologies that are consistent with observed data.
While here we have only explored the case of galaxy multi-tracer two-point functions; the multi-tracer bispectrum \cite{yamauchi_bispectrum_multitracer}, and cross-correlation with a matter tracer (e.g. via CMB lensing \cite{SchmittfullSeljak}) are potential directions in which to extend the methods presented here.

\acknowledgments
We thank Simone Ferraro, Noah Sailer, Boryana Hadzhiyska, and Anand Raichoor for helpful discussions about multitracer forecasts, the forecasts of Ref.~\cite{Noah21}, ELG-like and LRG-like IllustrisTNG galaxies, and observed DESI ELGs, respectively.
JMS is partially supported by a U.S. Dept. of Energy SCGSR award.
We thank the anonymous referee for useful and substantive comments on a draft version of this paper.

\appendix

\section{$F_{\fnlloc\fnlloc}$ expressions} 
\label{app:fisher_triple}

We first reproduce the expressions for the 2-tracer case, where $\mathbf{C}^{(2)} = 
    \frac{1}{V}\begin{pmatrix}
        \alpha^2 P_{2} +\frac{1}{\bar{n}_{1}} & r \alpha P_{2} \\
         &  P_{2} +\frac{1}{\bar{n}_{2}}
    \end{pmatrix}$, and add $F_{\alpha P_2}$ to the expressions for $F_{\alpha \alpha}$ and $F_{P_2 P_2}$ of Ref.~\cite{UrosMulti}\footnote{We find a slight difference here in the expression for $F_{P_{2} P_{2}}$ with respect to Ref.~\cite{UrosMulti}, though this does not affect any of the conclusions of Ref.~\cite{UrosMulti} since they are all stated in the sample-variance limit, where this difference disappears.}.

\begin{align*}
F_{\alpha \alpha} &= \frac{\alpha ^2 r^4+r^2 (X_{2}+1) \left(X_{1}-3 \alpha ^2\right)+2 \alpha ^2
   (X_{2}+1)^2}{\left(\alpha ^2 \left(-r^2+X_{2}+1\right)+X_{1}
   (X_{2}+1)\right)^2}\\
F_{P_2 P_2} &= \frac{2 \alpha ^2 X_{1} \left(r^2 (X_{2}-1)+1\right)+\alpha ^4 \left(2 r^4-2 r^2
   (X_{2}+2)+X_{2} (X_{2}+2)+2\right)+X_{1}^2}{2 \left(\alpha ^2
   P_{2} \left(-r^2+X_{2}+1\right)+P_{2} X_{1} (X_{2}+1)\right)^2}\\
F_{\alpha P_2} &= \frac{\alpha  r^2 X_{1} X_{2}+\alpha ^3
   \left(-r^2+X_{2}+1\right)^2}{P_{2} \left(\alpha ^2
   \left(-r^2+X_{2}+1\right)+X_{1} (X_{2}+1)\right)^2},
\end{align*}

where $X_{i} \equiv \frac{1}{\bar{n}_{i}P_{2}}$ quantifies the relative signal-to-noise.

In this work we used explicit symbolic expressions for the 3-tracer case, extending the expressions of Ref.~\cite{UrosMulti}.
The Fisher matrix element for $\fnlloc$ is given by the simple transformation:
\begin{equation}
    F_{\fnlloc \fnlloc} = \sum_{\lambda,\lambda'} \frac{\partial \lambda}{\partial \fnlloc} F_{\lambda \lambda'} \frac{\partial \lambda'}{\partial \fnlloc}
    \label{eqn:tensor_transform}
\end{equation}
where, in this work, $\lambda,\lambda' \in \{\alpha,\beta,P_{2}\}$.
This reduces to the two-tracer expressions when $\beta$ is neglected.

For the 3-tracer expressions, to make contact with the notation of Ref.~\cite{UrosMulti}, we can write eqn.~\ref{eqn:covariance} as 
\begin{equation}
\mathbf{C}^{(3)} = 
    \frac{1}{V}\begin{pmatrix}
        \alpha^2 P_{2} +\frac{1}{\bar{n}_{1}} & r_{12}\alpha P_{2} & r_{13} \alpha \beta P_{2}\\
         &  P_{2} +\frac{1}{\bar{n}_{2}} & r_{23}\beta P_{2}\\
         & & \beta^2 P_{2} +\frac{1}{\bar{n}_{3}}\\
    \end{pmatrix}
\end{equation}
The 3-tracer Fisher matrix element expressions for the multi-tracer parameters $\alpha,\beta,P_{2}$ are then given by:

\begin{align*}
    \begin{split}
    N_{\alpha \alpha} &= \alpha ^2 r_{12}^4 (\beta ^2+X_{3})^2-4 \alpha ^2 \beta ^2
   r_{12}^3 r_{13} r_{23} (\beta ^2+X_{3})\\
   &+r_{12}^2
   (\alpha ^2 (\beta ^4 (2 r_{13}^2 (2
   r_{23}^2+X_{2}+1)+3 (r_{23}^2-X_{2}-1))+\beta
   ^2 X_{3} (2 r_{13}^2 (X_{2}+1)+3 r_{23}^2-6
   (X_{2}+1))\\
   & -3 (X_{2}+1) X_{3}^2)+X_{1} (\beta
   ^2+X_{3}) (\beta ^2
   (-r_{23}^2+X_{2}+1)+(X_{2}+1) X_{3}))\\
   &-2 \beta
   ^2 r_{12} r_{13} r_{23} (\alpha ^2 (\beta ^2 (2
   r_{13}^2 (X_{2}+1)+3 (r_{23}^2-X_{2}-1))-3
   (X_{2}+1) X_{3})+\beta ^2 X_{1}
   (-r_{23}^2+X_{2}+1)\\
   &+X_{1} (X_{2}+1) X_{3})+\beta
   ^2 (X_{2}+1) X_{3} (r_{13}^2 (X_{2}+1) (X_{1}-3 \alpha
   ^2)+4 \alpha ^2 (-r_{23}^2+X_{2}+1))\\
   &+\beta ^4
   (\alpha ^2 r_{13}^4 (X_{2}+1)^2-r_{13}^2 (X_{2}+1)
   (r_{23}^2-X_{2}-1) (X_{1}-3 \alpha ^2)\\
   &+2 \alpha ^2
   (-r_{23}^2+X_{2}+1)^2)+2 \alpha ^2 (X_{2}+1)^2
   X_{3}^2\\
    D_{\alpha \alpha} &= (\alpha ^2 (X_{3}
   (-r_{12}^2+X_{2}+1)-\beta ^2 (r_{12}^2-2 r_{12}
   r_{13} r_{23}+(r_{13}^2-1)
   X_{2}+r_{13}^2+r_{23}^2-1))\\
   &+\beta ^2 X_{1}
   (-r_{23}^2+X_{2}+1)+X_{1} (X_{2}+1) X_{3})^2\\
    F_{\alpha \alpha}  &= \frac{N_{\alpha \alpha}}{D_{\alpha \alpha}}
    \end{split}
    \end{align*}

    \begin{align*}
    \begin{split}
    N_{\alpha \beta} &= \alpha  \beta  (-r_{12} r_{23}+r_{13} X_{2}+r_{13}) (\beta
   ^2 r_{13} (\alpha ^2 (r_{12}^2-2 r_{12} r_{13}
   r_{23}\\
   &+(r_{13}^2-1)
   X_{2}+r_{13}^2+r_{23}^2-1)+X_{1}
   (r_{23}^2-X_{2}-1))+\alpha ^2 r_{13} X_{3}
   (r_{12}^2-X_{2}-1)\\
   &+X_{1} X_{3} (-2 r_{12}
   r_{23}+r_{13} X_{2}+r_{13}))\\
    D_{\alpha \beta} &= (\alpha ^2 (X_{3}
   (-r_{12}^2+X_{2}+1)-\beta ^2 (r_{12}^2-2 r_{12}
   r_{13} r_{23}+(r_{13}^2-1)
   X_{2}+r_{13}^2+r_{23}^2-1))\\
   &+\beta ^2 X_{1}
   (-r_{23}^2+X_{2}+1)+X_{1} (X_{2}+1) X_{3})^2\\
    F_{\alpha \beta}  &= \frac{N_{\alpha \beta}}{D_{\alpha \beta}}
    \end{split}
    \end{align*}
    
    \begin{align*}
    \begin{split}
    N_{P_{2} \alpha} &= \alpha  (\alpha ^2 r_{12}^4 (\beta ^2+X_{3})^2-4 \alpha ^2
   \beta ^2 r_{12}^3 r_{13} r_{23} (\beta
   ^2+X_{3})+r_{12}^2 (2 \alpha ^2 (\beta ^4 (r_{13}^2
   (2 r_{23}^2+X_{2}+1)+r_{23}^2-X_{2}-1)\\
   &+\beta ^2
   X_{3} ((r_{13}^2-2)
   X_{2}+r_{13}^2+r_{23}^2-2)-(X_{2}+1)
   X_{3}^2)+X_{1} (\beta ^2 r_{23}^2 X_{3}+X_{2}
   (\beta ^2+X_{3})^2))\\
   &-2 \beta ^2 r_{12} r_{13}
   r_{23} (2 \alpha ^2 (\beta ^2 ((r_{13}^2-1)
   X_{2}+r_{13}^2+r_{23}^2-1)-(X_{2}+1) X_{3})+X_{1}
   (\beta ^2 X_{2}+2 X_{2} X_{3}+X_{3}))\\
   &+\beta ^4
   (r_{13}^2 r_{23}^2 X_{1} X_{2}+\alpha ^2 (r_{13}^2
   (X_{2}+1)+r_{23}^2-X_{2}-1)^2)+\beta ^2 (X_{2}+1)
   X_{3} (r_{13}^2 (X_{2}+1) (X_{1}-2 \alpha ^2)\\
   &+2 \alpha
   ^2 (-r_{23}^2+X_{2}+1))+\alpha ^2 (X_{2}+1)^2
   X_{3}^2)\\
    D_{P_{2}\alpha} &= P_{2} (\alpha ^2 (X_{3}
   (-r_{12}^2+X_{2}+1)-\beta ^2 (r_{12}^2-2 r_{12}
   r_{13} r_{23}+(r_{13}^2-1)
   X_{2}+r_{13}^2+r_{23}^2-1))\\
   &+\beta ^2 X_{1}
   (-r_{23}^2+X_{2}+1)+X_{1} (X_{2}+1) X_{3})^2\\
    F_{P_{2} \alpha}  &= \frac{N_{P_{2} \alpha}}{D_{P_{2}\alpha}}
    \end{split}
    \end{align*}
    
    \begin{align*}
    \begin{split}
    N_{P_{2}P_{2}} &= 2 \alpha^2 X_{1} (\beta^4 (2 (r_{23}^2-1)
   (r_{12}^2-2 r_{12} r_{13}
   r_{23}+r_{13}^2+r_{23}^2-1)\\
   &+2 (r_{13}^2-1)
   (r_{23}^2-1) X_{2}-(r_{13}^2-1)
   X_{2}^2)+\beta ^2 X_{3} (2 r_{12}^2
   (r_{23}^2+X_{2}-1)\\
   &-6 r_{12} r_{13} r_{23}
   X_{2}+r_{13}^2 X_{2} (X_{2}+2)+2 r_{23}^2
   (X_{2}-1)+2)+X_{3}^2 (r_{12}^2
   (X_{2}-1)+1))\\
   &+  \alpha ^4 (\beta ^4 (4
   (r_{13}^2-1) X_{2} (r_{12}^2-2 r_{12} r_{13}
   r_{23}+r_{13}^2+r_{23}^2-1)+3 (r_{12}^2-2 r_{12}
   r_{13} r_{23}+r_{13}^2+r_{23}^2-1)^2\\
   &+2(r_{13}^2-1)^2 X_{2}^2)+2 \beta ^2 X_{3} (2
   (r_{12}^2-1) (r_{12}^2-2 r_{12} r_{13}
   r_{23}+r_{13}^2+r_{23}^2-1)+2 (r_{12}^2-1)
   (r_{13}^2-1) X_{2}\\
   &-(r_{13}^2-1)
   X_{2}^2)+X_{3}^2 (2 r_{12}^4-2 r_{12}^2
   (X_{2}+2)+X_{2} (X_{2}+2)+2))\\
   &+X_{1}^2 (2 \beta^2
   X_{3} (r_{23}^2 (X_{2}-1)+1)+\beta ^4 (2 r_{23}^4-2
   r_{23}^2 (X_{2}+2)+X_{2} (X_{2}+2)+2)+X_{3}^2)\\
   D_{P_{2}P_{2}} &= 2
   P_{2}^2 (\alpha ^2 (X_{3}
   (-r_{12}^2+X_{2}+1)-\beta^2 (r_{12}^2-2 r_{12}
   r_{13} r_{23}+(r_{13}^2-1)
   X_{2}+r_{13}^2+r_{23}^2-1))+\beta^2 X_{1} \\
   &~ (-r_{23}^2+X_{2}+1)+X_{1} (X_{2}+1) X_{3})^2\\
    F_{P_{2}P_{2}}  &= \frac{N_{P_{2}P_{2}}}{D_{P_{2}P_{2}}}
    \end{split}
\end{align*}

We do not write the expressions for $F_{\beta P_{2}}$ or $F_{\beta \beta}$ since these are the same as $F_{\alpha P_2}, F_{\alpha \alpha}$ with interchanged arguments $\beta \longleftrightarrow \alpha$.

The Jacobian factors for the transformation in eqn.~\ref{eqn:tensor_transform} for the parameters $\alpha$ and $\beta$ are:
\begin{align}
    \frac{\partial \alpha}{\partial \fnlloc} & = \alpha\left(\frac{b_{\phi,1}}{b_{1}} - \frac{b_{\phi,2}}{b_{2}} \right)\mathcal{M}^{-1}\\
    \frac{\partial P_2}{\partial \fnlloc} & = 2 b_{\phi,2} \mathcal{M}^{-1} \frac{P_{2}}{b_{2}}
\end{align}

where we dropped the $k$-dependence of $\mathcal{M}$ and $P_{2}$ to be consistent with the previous expressions, and again the $\beta$ expression is analogous to the $\alpha$ expression.

In the sample variance limit, at leading order in $X_{i}$, and assuming $b_{1}=b_{2}=b_{3}$, the ratio of the $\fnlloc$ Fisher information is:

\begin{align}
\label{eqn:fish_info_23_rat_1}
    \frac{
    F^{(3)}_{\fnlloc}}{F^{(2)}_{\fnlloc\fnlloc}}&=
\frac{2 \left((b_{\phi,1}- b_{\phi,2})^{2} +(b_{\phi,2}- b_{\phi,3})^{2} +(b_{\phi,1}- b_{\phi,3})^{2}\right)
}{3 (b_{\phi,1}-
   b_{\phi,2})^2}\\
   & \quad-\frac{2 X_{2} \left(43 b_{\phi,1}^2+59 b_{\phi,1}b_{\phi,2}-145 b_{\phi,1} b_{\phi,3}-59 b_{\phi,2}^2+59 b_{\phi,2} b_{\phi,3}+43 b_{\phi,3}^2\right)}{27
   (b_{\phi,1}-b_{\phi,2})^2}\nonumber
\end{align}

\begin{equation}
    \begin{split}
    \frac{F^{(3,P)}_{\fnlloc\fnlloc}}{F^{(2,P)}_{\fnlloc\fnlloc}}&= \frac{F^{(3)}_{\fnlloc}}{F^{(2)}_{\fnlloc\fnlloc}}-\frac{4 b_{\phi,2} X_2}{9 (b_{\phi,1}-b_{\phi,2})^4} [18 b_{\phi,1}^3+b_{\phi,1}^2 (33 b_{\phi,2}-86 b_{\phi,3})\\ &\quad\quad\quad\quad\quad\quad\quad\quad\quad\quad\quad\quad\quad+4 b_{\phi,1} \left(-4 b_{\phi,2}^2+4 b_{\phi,2} b_{\phi,3}+13 b_{\phi,3}^2\right)\\&\quad\quad\quad\quad\quad\quad\quad\quad\quad\quad\quad\quad\quad+17 b_{\phi,2}^2 (b_{\phi,2}-2 b_{\phi,3})]
    \end{split}
\label{eqn:fish_info_23_rat_2}
\end{equation}
when considering only relative amplitude parameters, and when including $P_{2}$, respectively.
These expressions illustrate that these contributions both approach the first line of eqn.~\ref{eqn:fish_info_23_rat_1} as $X_{2}\to 0$, but can be significantly different otherwise.

Similarly, when considering the ratio of the Fisher information when using $P_{2}$ as a parameter or not for two tracers, we have:
\begin{equation}
    \frac{F^{(2,P)}_{\fnlloc}}{F^{(2)}_{\fnlloc\fnlloc}} = 1 + X_{2}\frac{52 b_{\phi,1} b_{\phi,2}}{3 (b_{\phi,1} - b_{\phi,2})^2}.
\end{equation}

Again, we see that in the sample variance limit the Fisher information is equivalent, but outside of the case where $X_{2}\to 0$ the exact values of $\bphi$ in each sample can drive the ratio above or below one (e.g. in the latter case if the values of $b_{\phi,1}, b_{\phi,2}$ have different signs).

We now make a few brief comments about these expressions, which may  be useful for the interested reader of the main text.
To be consistent with the notation of Ref.~\cite{UrosMulti} we have taken the two-tracer case to have LPNG biases $b_{\phi,1}$ and $b_{\phi,2}$, so care should be taken to interpret these expressions appropriately (and not to simply associate $b_{\phi,1}$ with $b_{\phi}^{-}$ or $b_{\phi,3}$ with $b_{\phi}^{+}$).
In Section~\ref{subsec:fisher_results}, for future surveys we considered a symmetric $\Delta p$ for the upper and lower tertiles, though in general this will not be the case.
Eqn.~\ref{eqn:fish_info_23_rat_1} also indicates that the exact departure from this symmetry may affect whether a 2-tracer of 3-tracer forecast is desirable.
We also note that, in general, both the linear bias $b$ of the sample and the number density $\bar{n}$ will spoil these statements as we leave the sample variance limit and $b$ changes over the splits.

\bibliographystyle{JHEP}
\bibliography{fnlbphitng.bib}

\end{document}